\documentclass[twocolumn, twocolappendix]{aastex631}

\usepackage[hyphens]{url}

\shorttitle{Circumnuclear gas from ECLE  AT\,2022upj}
\shortauthors{Newsome et al.}

\graphicspath{{./}}
\usepackage{amsmath}
\begin{document}


\title{Mapping the Inner 0.1 pc of a Supermassive Black Hole Environment with the Tidal Disruption Event and Extreme Coronal Line Emitter AT\,2022upj}

\newcommand{\LCO}{\affiliation{Las Cumbres Observatory, 6740 Cortona Drive, Suite 102, Goleta, CA 93117-5575, USA}}
\newcommand{\UCSB}{\affiliation{Department of Physics, University of California, Santa Barbara, CA 93106-9530, USA}}
\newcommand{\Iair}
{\affiliation{CIFAR Azrieli Global Scholars program, CIFAR, Toronto, Canada}}
\newcommand{\KITP}{\affiliation{Kavli Institute for Theoretical Physics, University of California, Santa Barbara, CA 93106-4030, USA}}
\newcommand{\UCD}{\affiliation{Department of Physics and Astronomy, University of California, Davis, 1 Shields Avenue, Davis, CA 95616-5270, USA}}
\newcommand{\WIS}{\affiliation{Department of Particle Physics and Astrophysics, Weizmann Institute of Science, 76100 Rehovot, Israel}}
\newcommand{\OKC}{\affiliation{Oskar Klein Centre, Department of Astronomy, Stockholm University, Albanova University Centre, SE-106 91 Stockholm, Sweden}}
\newcommand{\OAPD}{\affiliation{INAF -- Osservatorio Astronomico di Padova, Vicolo dell'Osservatorio 5, I-35122 Padova, Italy}}
\newcommand{\Caltech}{\affiliation{Cahill Center for Astronomy and Astrophysics, California Institute of Technology, Mail Code 249-17, Pasadena, CA 91125, USA}}
\newcommand{\GSFC}{\affiliation{Astrophysics Science Division, NASA Goddard Space Flight Center, Mail Code 661, Greenbelt, MD 20771, USA}}
\newcommand{\UMD}{\affiliation{Joint Space-Science Institute, University of Maryland, College Park, MD 20742, USA}}
\newcommand{\UCB}{\affiliation{Department of Astronomy, University of California, Berkeley, CA 94720-3411, USA}}
\newcommand{\TTU}{\affiliation{Department of Physics, Texas Tech University, Box 41051, Lubbock, TX 79409-1051, USA}}
\newcommand{\STScI}{\affiliation{Space Telescope Science Institute, 3700 San Martin Drive, Baltimore, MD 21218-2410, USA}}
\newcommand{\UT}{\affiliation{University of Texas at Austin, 1 University Station C1400, Austin, TX 78712-0259, USA}}
\newcommand{\IoA}{\affiliation{Institute of Astronomy, University of Cambridge, Madingley Road, Cambridge CB3 0HA, UK}}
\newcommand{\QUB}{\affiliation{Astrophysics Research Centre, School of Mathematics and Physics, Queen's University Belfast, Belfast BT7 1NN, UK}}
\newcommand{\IPAC}{\affiliation{Spitzer Science Center, California Institute of Technology, Pasadena, CA 91125, USA}}
\newcommand{\JPL}{\affiliation{Jet Propulsion Laboratory, California Institute of Technology, 4800 Oak Grove Dr, Pasadena, CA 91109, USA}}
\newcommand{\Southampton}{\affiliation{Department of Physics and Astronomy, University of Southampton, Southampton SO17 1BJ, UK}}
\newcommand{\LANL}{\affiliation{Space and Remote Sensing, MS B244, Los Alamos National Laboratory, Los Alamos, NM 87545, USA}}
\newcommand{\Tsinghua}{\affiliation{Physics Department and Tsinghua Center for Astrophysics, Tsinghua University, Beijing, 100084, People's Republic of China}}
\newcommand{\NAOC}{\affiliation{National Astronomical Observatory of China, Chinese Academy of Sciences, Beijing, 100012, People's Republic of China}}
\newcommand{\Itagaki}{\affiliation{Itagaki Astronomical Observatory, Yamagata 990-2492, Japan}}
\newcommand{\Einstein}{\altaffiliation{Einstein Fellow}}
\newcommand{\Hubble}{\altaffiliation{Hubble Fellow}}
\newcommand{\CfA}{\affiliation{Center for Astrophysics \textbar{} Harvard \& Smithsonian, 60 Garden Street, Cambridge, MA 02138-1516, USA}}
\newcommand{\UA}{\affiliation{Steward Observatory, University of Arizona, 933 North Cherry Avenue, Tucson, AZ 85721-0065, USA}}
\newcommand{\MPIA}{\affiliation{Max-Planck-Institut f\"ur Astrophysik, Karl-Schwarzschild-Stra\ss{}e 1, D-85748 Garching, Germany}}
\newcommand{\DSFP}{\altaffiliation{LSSTC Data Science Fellow}}
\newcommand{\HCO}{\affiliation{Harvard College Observatory, 60 Garden Street, Cambridge, MA 02138-1516, USA}}
\newcommand{\Carnegie}{\affiliation{Observatories of the Carnegie Institute for Science, 813 Santa Barbara Street, Pasadena, CA 91101-1232, USA}}
\newcommand{\TAU}{\affiliation{School of Physics and Astronomy, Tel Aviv University, Tel Aviv 69978, Israel}}
\newcommand{\Edinburgh}{\affiliation{Institute for Astronomy, University of Edinburgh, Royal Observatory, Blackford Hill EH9 3HJ, UK}}
\newcommand{\Birmingham}{\affiliation{Birmingham Institute for Gravitational Wave Astronomy and School of Physics and Astronomy, University of Birmingham, Birmingham B15 2TT, UK}}
\newcommand{\Bath}{\affiliation{Department of Physics, University of Bath, Claverton Down, Bath BA2 7AY, UK}}
\newcommand{\CTIO}{\affiliation{Cerro Tololo Inter-American Observatory, National Optical Astronomy Observatory, Casilla 603, La Serena, Chile}}
\newcommand{\Potsdam}{\affiliation{Institut f\"ur Physik und Astronomie, Universit\"at Potsdam, Haus 28, Karl-Liebknecht-Str. 24/25, D-14476 Potsdam-Golm, Germany}}
\newcommand{\INPE}{\affiliation{Instituto Nacional de Pesquisas Espaciais, Avenida dos Astronautas 1758, 12227-010, S\~ao Jos\'e dos Campos -- SP, Brazil}}
\newcommand{\UNC}{\affiliation{Department of Physics and Astronomy, University of North Carolina, 120 East Cameron Avenue, Chapel Hill, NC 27599, USA}}
\newcommand{\Ohio}{\affiliation{Astrophysical Institute, Department of Physics and Astronomy, 251B Clippinger Lab, Ohio University, Athens, OH 45701-2942, USA}}
\newcommand{\AAS}{\affiliation{American Astronomical Society, 1667 K~Street NW, Suite 800, Washington, DC 20006-1681, USA}}
\newcommand{\MMT}{\affiliation{MMT and Steward Observatories, University of Arizona, 933 North Cherry Avenue, Tucson, AZ 85721-0065, USA}}
\newcommand{\Geneva}{\affiliation{ISDC, Department of Astronomy, University of Geneva, Chemin d'\'Ecogia, 16 CH-1290 Versoix, Switzerland}}
\newcommand{\IUCAA}{\affiliation{Inter-University Center for Astronomy and Astrophysics, Post Bag 4, Ganeshkhind, Pune, Maharashtra 411007, India}}
\newcommand{\CMU}{\affiliation{Department of Physics, Carnegie Mellon University, 5000 Forbes Avenue, Pittsburgh, PA 15213-3815, USA}}
\newcommand{\NAOJ}{\affiliation{Division of Science, National Astronomical Observatory of Japan, 2-21-1 Osawa, Mitaka, Tokyo 181-8588, Japan}}
\newcommand{\IfA}{\affiliation{Institute for Astronomy, University of Hawai`i, 2680 Woodlawn Drive, Honolulu, HI 96822-1839, USA}}
\newcommand{\UCSC}{\affiliation{Department of Astronomy and Astrophysics, University of California, Santa Cruz, CA 95064-1077, USA}}
\newcommand{\Purdue}{\affiliation{Department of Physics and Astronomy, Purdue University, 525 Northwestern Avenue, West Lafayette, IN 47907-2036, USA}}
\newcommand{\Princeton}{\affiliation{Department of Astrophysical Sciences, Princeton University, 4 Ivy Lane, Princeton, NJ 08540-7219, USA}}
\newcommand{\Moore}{\affiliation{Gordon and Betty Moore Foundation, 1661 Page Mill Road, Palo Alto, CA 94304-1209, USA}}
\newcommand{\Durham}{\affiliation{Department of Physics, Durham University, South Road, Durham, DH1 3LE, UK}}
\newcommand{\JHU}{\affiliation{Department of Physics and Astronomy, The Johns Hopkins University, 3400 North Charles Street, Baltimore, MD 21218, USA}}
\newcommand{\Toronto}{\affiliation{David A.\ Dunlap Department of Astronomy and Astrophysics, University of Toronto,\\ 50 St.\ George Street, Toronto, Ontario, M5S 3H4 Canada}}
\newcommand{\Duke}{\affiliation{Department of Physics, Duke University, Campus Box 90305, Durham, NC 27708, USA}}
\newcommand{\NCU}{\affiliation{Graduate Institute of Astronomy, National Central University, 300 Jhongda Road, 32001 Jhongli, Taiwan}}
\newcommand{\Columbia}{\affiliation{Department of Physics and Columbia Astrophysics Laboratory, Columbia University, Pupin Hall, New York, NY 10027, USA}}
\newcommand{\Flatiron}{\affiliation{Center for Computational Astrophysics, Flatiron Institute, 162 5th Avenue, New York, NY 10010-5902, USA}}
\newcommand{\CIERA}{\affiliation{Center for Interdisciplinary Exploration and Research in Astrophysics and Department of Physics and Astronomy, \\Northwestern University, 1800 Sherman Avenue, 8th Floor, Evanston, IL 60201, USA}}
\newcommand{\GeminiObs}{\affiliation{Gemini Observatory, 670 North A`ohoku Place, Hilo, HI 96720-2700, USA}}
\newcommand{\Keck}{\affiliation{W.~M.~Keck Observatory, 65-1120 M\=amalahoa Highway, Kamuela, HI 96743-8431, USA}}
\newcommand{\UW}{\affiliation{Department of Astronomy, University of Washington, 3910 15th Avenue NE, Seattle, WA 98195-0002, USA}}
\newcommand{\DiRAC}{\altaffiliation{DiRAC Fellow}}
\newcommand{\USask}{\affiliation{Department of Physics \& Engineering Physics, University of Saskatchewan, 116 Science Place, Saskatoon, SK S7N 5E2, Canada}}
\newcommand{\Thacher}{\affiliation{Thacher School, 5025 Thacher Road, Ojai, CA 93023-8304, USA}}
\newcommand{\Rutgers}{\affiliation{Department of Physics and Astronomy, Rutgers, the State University of New Jersey,\\136 Frelinghuysen Road, Piscataway, NJ 08854-8019, USA}}
\newcommand{\FSU}{\affiliation{Department of Physics, Florida State University, 77 Chieftan Way, Tallahassee, FL 32306-4350, USA}}
\newcommand{\Melbourne}{\affiliation{School of Physics, The University of Melbourne, Parkville, VIC 3010, Australia}}
\newcommand{\ASTROthreeD}{\affiliation{ARC Centre of Excellence for All Sky Astrophysics in 3 Dimensions (ASTRO 3D)}}
\newcommand{\Stromlo}{\affiliation{Mt.\ Stromlo Observatory, The Research School of Astronomy and Astrophysics, Australian National University, ACT 2601, Australia}}
\newcommand{\NCPAS}{\affiliation{National Centre for the Public Awareness of Science, Australian National University, Canberra, ACT 2611, Australia}}
\newcommand{\TAMU}{\affiliation{Department of Physics and Astronomy, Texas A\&M University, 4242 TAMU, College Station, TX 77843, USA}}
\newcommand{\Mitchell}{\affiliation{George P.\ and Cynthia Woods Mitchell Institute for Fundamental Physics \& Astronomy, College Station, TX 77843, USA}}
\newcommand{\ESO}{\affiliation{European Southern Observatory, Alonso de C\'ordova 3107, Casilla 19, Santiago, Chile}}
\newcommand{\ICE}{\affiliation{Institute of Space Sciences (ICE, CSIC), Campus UAB, Carrer
de Can Magrans, s/n, E-08193 Barcelona, Spain}}
\newcommand{\IEEC}{\affiliation{Institut d'Estudis Espacials de Catalunya, Gran Capit\`a, 2-4, Edifici Nexus, Desp.\ 201, E-08034 Barcelona, Spain}}
\newcommand{\Warwick}{\affiliation{Department of Physics, University of Warwick, Gibbet Hill Road, Coventry CV4 7AL, UK}}
\newcommand{\Macquarie}{\affiliation{School of Mathematical and Physical Sciences, Macquarie University, NSW 2109, Australia}}
\newcommand{\AAARC}{\affiliation{Astronomy, Astrophysics and Astrophotonics Research Centre, Macquarie University, Sydney, NSW 2109, Australia}}
\newcommand{\Capodimonte}{\affiliation{INAF -- Capodimonte Astronomical Observatory, Salita Moiariello 16, I-80131 Napoli, Italy}}
\newcommand{\INFNNapoli}{\affiliation{INFN -- Napoli, Strada Comunale Cinthia, I-80126 Napoli, Italy}}
\newcommand{\ICRANet}{\affiliation{ICRANet, Piazza della Repubblica 10, I-65122 Pescara, Italy}}
\newcommand{\UVA}{\affiliation{Department of Astronomy, University of Virginia, Charlottesville, VA 22904, USA}}

\author[0000-0001-9570-0584]{Megan Newsome}
\LCO\UCSB
\author[0000-0001-7090-4898]{Iair Arcavi}
\TAU
\author[0000-0003-4253-656X]{D. Andrew Howell}
\LCO\UCSB
\author[0000-0001-5807-7893]{Curtis McCully}
\LCO
\author[0000-0003-0794-5982]{Giacomo Terreran}
\LCO
\author[0000-0002-0832-2974]{Griffin Hosseinzadeh}
\UA
\author[0000-0002-4924-444X]{K. Azalee Bostroem}
\UA
\altaffiliation{LSSTC Catalyst Fellow}
\author[0000-0002-7579-1105]{Yael Dgany}
\TAU
\author[0000-0003-4914-5625]{Joseph Farah}
\LCO\UCSB
\author[0009-0007-8485-1281]{Sara Faris}
\TAU
\author[0000-0003-0209-9246]{Estefania Padilla-Gonzalez}
\LCO\UCSB
\author[0000-0002-7472-1279]{Craig Pellegrino}
\UVA
\author[0000-0002-1895-6639]{Moira Andrews}
\LCO\UCSB

 
\begin{abstract}

Extreme coronal line emitters (ECLEs) are objects showing transient high-ionization lines in the centers of galaxies. They have been attributed to echoes of high-energy flares of ionizing radiation, such as those produced by tidal disruption events (TDEs), but have only recently been observed within hundreds of days after an optical transient was detected. AT\,2022upj is a nuclear UV-optical flare at $z=0.054$ with spectra showing $[$\ion{Fe}{10}$]$ $\lambda$6375 and $[$\ion{Fe}{14}$]$ $\lambda$5303 during the optical peak, the earliest presence of extreme coronal lines during an ongoing transient. AT\,2022upj is also the second ever ECLE (and first with a concurrent flare) to show broad He II $\lambda$4686 emission, a key signature of optical/UV TDEs. We also detect X-ray emission during the optical transient phase, which may be related to the source of ionizing photons for the extreme coronal lines. Finally, we analyze the spectroscopic evolution of each emission line and find that $[$\ion{Fe}{10}$]$ and $[$\ion{Fe}{14}$]$ weaken within 400d of optical peak, while $[$\ion{Fe}{7}$]$ $\lambda$5720, $[$\ion{Fe}{7}$]$ $\lambda6087$, and $[$\ion{O}{3}$]$ $\lambda$$\lambda$4959,5007 emerge over the same period. The velocities of the iron lines indicate circumnuclear gas within 0.1pc of the central supermassive black hole (SMBH), while a dust echo inferred from NEOWISE data indicates that circumnuclear dust lies at a minimum of 0.4pc away, providing evidence of stratified material around a SMBH. AT\,2022upj is thus the first confirmed ECLE-TDE with clear signatures of both classes and with spectroscopic evolution on a $\sim$year-long timescale. This event helps unveil the impact of highly energetic flares such as TDEs on the complex environments around SMBHs.

\end{abstract}

\section{Introduction} \label{sec:intro}

Supermassive black holes (SMBHs) at the centers of galaxies produce tidal forces that can disrupt a star's self-gravity upon close contact \citep{1975Natur.254..295H, Rees1988}. These tidal disruption events (TDEs) create a stream of stellar material that accretes onto the SMBH, producing a flare in the X-ray, ultraviolet, and/or optical bands that can outshine the host galaxy. Observations of TDEs thus offer a window into the otherwise quiescent SMBH population, and can teach us about accretion physics, jet formation and the environments around SMBHs. The discovery rate of TDEs is rapidly increasing \citep[e.g.][]{Graham2019, Gezari2021}, with over 60 events reported to the Transient Name Server\footnote{\url{https://www.wis-tns.org/}} in the last five years. With these larger samples, we now see groupings of TDE types by spectral signatures, with most showing broad emission of only H$\alpha$ $\lambda$6563, only He II $\lambda$4686, or both \citep[e.g.][]{Gezari2012, Gezari2021, VanVelzen21, Nicholl2022, Hammerstein2023}. 

\begin{figure*}[t!]
    \centering
    \includegraphics[scale=0.7]{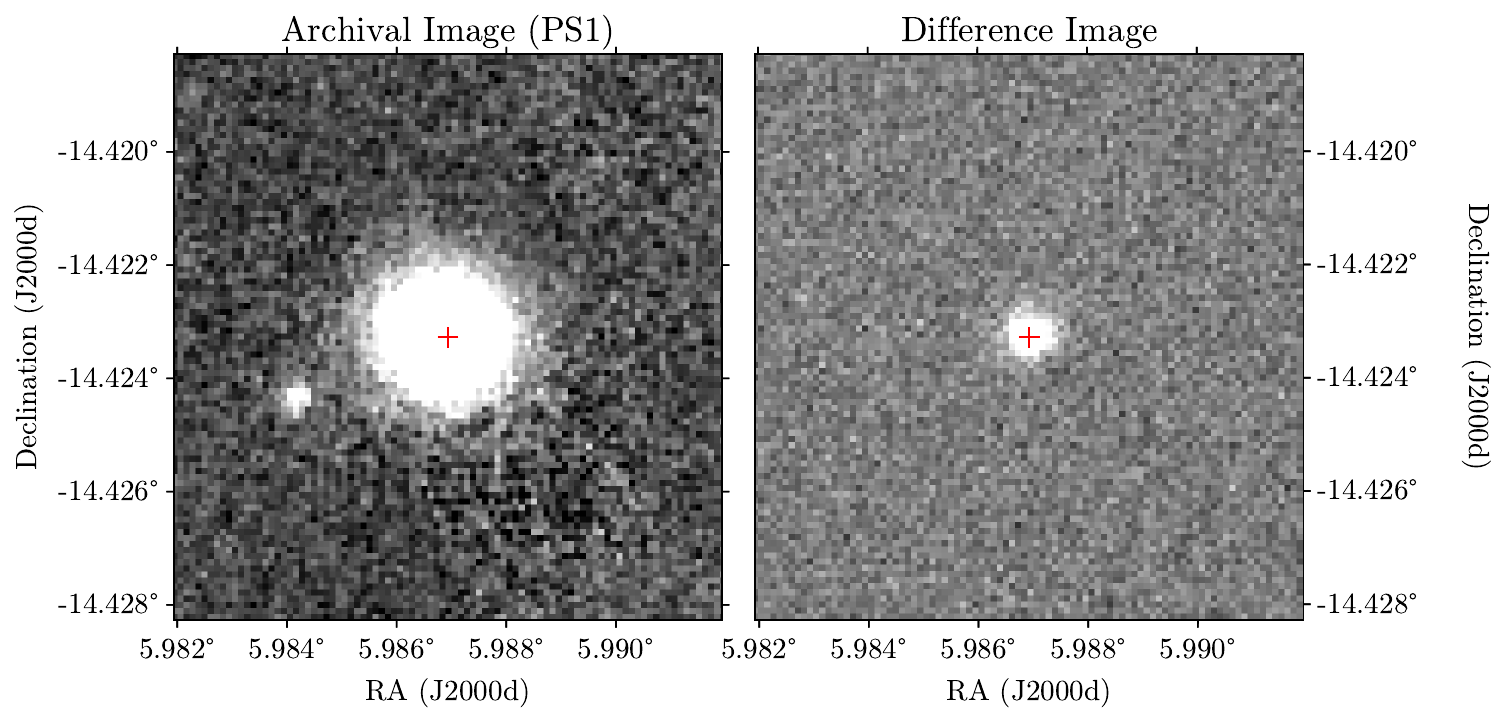}
    \caption{Archival image of host galaxy WISEA J002356.88-142523.9 from Pan-STARRS1 in $g$-band (left) compared with the location of the transient as observed in subtracted $g$-band images with Las Cumbres Observatory (right). The transient is located within $0.1''$ (0.1 kpc) from the host's center at (RA, Dec) = (00:23:56.9, $-$14:25:24) as marked by the red crossbar.}
    \label{fig:nuclear_placement}
\end{figure*}

Another type of transient phenomenon associated with SMBHs, extreme coronal line emitters (ECLEs), was initially identified in seven galaxies which showed nuclear emission of highly ionized $[$\ion{Fe}{10}$]$ $\lambda$6375 and $[$\ion{Fe}{14}$]$ $\lambda$5303 \citep{Komossa2008, Wang2011, Wang2012, Yang2013}. Despite having no simultaneous direct evidence of tidal disruption flares, these spectral features were suggested to result from an extreme ultraviolet to soft X-ray outburst of radiation associated with a TDE. 

\cite{Clark2023} followed these early-identified ECLEs and confirmed five as single transient events likely to be caused by a TDE due to their lack of recurring coronal lines. Ten recent ECLEs have been found to show the emergence of coronal lines after the detection of an optical brightening in a nuclear source \citep[][Clark et al. in prep]{Neustadt2020, Onori2022, Hinkle2023, 2023RAA....23b5012L, Short2023, Somalwar2023, Yao2023, Koljonen2024, 2024ApJ...966..136W}, which were all reported to also be TDEs, but in each case the coronal lines were not detected until dozens to hundreds of days after optical peak, as listed in Table \ref{table:litreview}. From this TDE-ECLE sample only AT 2017gge \citep{Onori2022} and AT 2019qiz \citep{Short2023} exhibited X-ray activity prior to, or concurrent with, the coronal lines. For the other five objects, the ionizing X-ray source was either not detected at all, or only detected after the emergence and disappearance of coronal iron lines, which may mean X-ray emission was previously obscured in these cases. 

ECLEs can thus be used as indicators of obscured TDEs. Indeed, \cite{Hinkle2023} made the first statistical argument for a connection between ECLEs and TDEs via host property analysis of all known ECLEs. Furthermore, ECLEs can serve as probes of the circumnuclear gas around SMBHs, which is otherwise inaccessible observationally.
\begin{deluxetable}{cccc}[t]\label{table:litreview}
    \caption{Tidal disruption events that were later discovered to show extreme coronal line emission, where the Phase column shows the number of days after discovery that the coronal lines were measured. No nuclear transient has been discovered with coronal iron lines that were concurrent with the optical peak of the flare before AT 2022upj. We also list the transients for which $g$ or $c$-band forced photometry exists with magnitude errors $<0.25$ mag beyond +200d from optical peak in the fourth column, used to compare light curve evolution with AT 2022upj.}
    \tablehead{
    \colhead{Name} & \colhead{Phase} & \colhead{Source} & 
    {+200d $g/c$?}\\
    \colhead{} & \colhead{(days)} & \colhead{} & 
    {}}
    \startdata
        AT 2017gge & +200 & \cite{Onori2022} & No\\
        AT 2018bcb & +35 & \cite{Neustadt2020} & Yes\\
        AT 2018gn & +1528 & \cite{2024ApJ...966..136W} & Yes \\
        AT 2018dyk & \nodata & Clark et al. (in prep.) & Yes\\
        AT 2019qiz & +428 & \cite{Short2023} & No\\
        AT 2020vdq & +529 & \cite{Somalwar2023} & No\\
        AT 2021dms & +175 & \cite{Hinkle2023} & Yes\\
        AT 2021qth & +300 & \cite{Yao2023} & No\\
        AT 2021acak & +170 & \cite{2023RAA....23b5012L} & Yes \\
        AT 2022fpx & +11 & \cite{Koljonen2024} & Yes\\
        AT 2022upj & 0 & This work & Yes\\
    \enddata
\end{deluxetable}

AT 2022upj was discovered by the Zwicky Transient Facility (ZTF) on MJD 59822.39 (August 31 2022) as a nuclear transient \citep{2022TNSTR2690....1F}. Its location in the center of the galaxy WISEA J002356.88-142523.9 at (RA$_{\text{J2000}}$, Dec$_{\text{J2000}}$) = (00:23:56.86, $-$14:25:23.4), as shown in Figure \ref{fig:nuclear_placement}, and blue color ($g-r = -0.068$) upon first available multi-band detection (MJD = 59846.3, from ZTF) indicated a possible TDE origin, after which we began our ground-based optical follow-up with Las Cumbres Observatory \citep{2013PASP..125.1031B}.  Spectroscopy from Las Cumbres Observatory's Faulkes Telescope South revealed the bright and narrow high-ionization emission lines $[$\ion{Fe}{10}$]$$\lambda6375$ and $[$\ion{Fe}{14}$]$ $\lambda$5303, each matched to a redshift of $z=0.054$, as early as MJD 59885.0, classifying AT 2022upj as an ECLE \citep{2022TNSAN.236....1N, 2022TNSCR3231....1N}. An additional broad He II $\lambda$4686 emission feature, typical of TDEs is also present, seen only once before in SDSS 0748+4712 \citep{Wang2011}. 

AT 2022upj is the first nuclear transient to show coronal lines concurrent with the optical peak of the light curve. Subsequent observations with the Niel Gehrels Swift Observatory (Swift) also make AT 2022upj the first ECLE to show concurrent soft X-ray emission alongside the transient peak and coronal lines, providing potential evidence of the high-energy ionizing source.

\begin{figure*}[t!]
    \centering
    \includegraphics[]{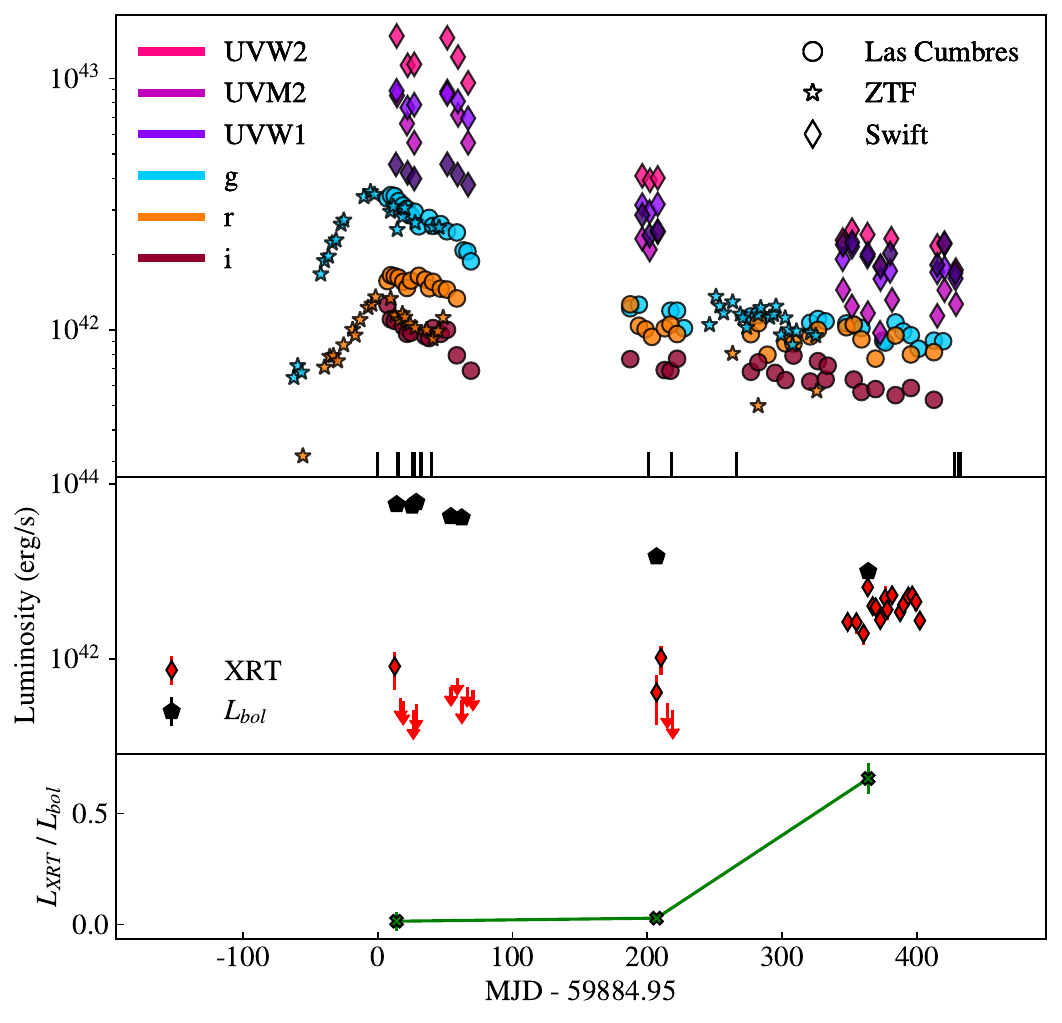}
    \caption{Top: Host-subtracted and Galactic-extinction corrected optical and UV light curves of AT\,2022upj as observed by Las Cumbres Observatory, ZTF, and Swift. The light curves in each filter are offset from one another for clarity. The $gri$ points from Las Cumbres were subtracted with PS1 archival templates. Vertical black tick marks indicate when spectroscopy was simultaneously obtained. Middle: AT\,2022upj X-ray (0.1--10keV) luminosity from Swift XRT (red diamonds or downward arrows for $3\sigma$ upper limits) compared to the pseudo-bolometric luminosity, calculated from blackbody fits to the Swift UVOT and Las Cumbres + ZTF optical photometry (black pentagons). Bottom: The ratio of $L_{\text{XRT}}$ to $L_{\text{bol}}$ (green x-marks). The observed X-ray output is less than $2\%$ of the UV-optical pseudo-bolometric luminosity until after +300d from optical peak. The data behind this figure are available as a machine-readable table.}
    \label{fig:offset_lightcurve_all}
\end{figure*}

Here we present the UV, optical, X-ray, and mid-IR light curve as well as the optical spectroscopic evolution of AT 2022upj from continued follow-up via Las Cumbres Observatory, ZTF, Swift, and the Wide-field Infrared Survey Explorer \citep[WISE;][]{2010AJ....140.1868W}. We review our methods of data extraction and reduction in \S\ref{sec:style}, review the results of the data including coronal line flux evolution and line ratio evolution in \S \ref{sec:results}, discuss the implications of the coronal line observations on the properties of the circumnuclear material including proximity to the central SMBH and relationship to host evolution in \S\ref{sec:gas}, and summarize the novelty of the short-term ECLE variation in \S\ref{sec:conclusion}. Throughout this work we adopt the \cite{2020A&A...641A...6P} cosmology with H$_0$ = 67.7 km s$^{-1}$ Mpc$^{-1}$.

\section{Observations and Data Processing} \label{sec:style}

Throughout this work, we use observations taken by Las Cumbres Observatory, ZTF, Swift, and WISE. We correct all photometry for Galactic extinction with $A_V=0.083$ mag \citep[][]{2011ApJ...737..103S} using the Calzetti dust law \citep[][]{Calzetti2000}. 

\subsection{Las Cumbres and ZTF Photometry}

We obtained $gri$-band images with the Sinistro cameras on Las Cumbres' 1.0m telescopes, performed image subtraction, and extracted point-spread-function (PSF) photometry with the \texttt{lcogtsnpipe} pipeline\footnote{\url{https://github.com/LCOGT/lcogtsnpipe}} \citep{Valenti16}. Zeropoints for all bands were calculated from the magnitudes of field stars as listed in the AAVSO Photometry All-Sky Survey \citep{2009AAS...21440702H}. 
Archival images from the Panoramic Survey Telescope and Rapid Response System, Pan-STARRS1 \citep[PS1;][]{2010SPIE.7733E..0EK}, were used as templates for the image subtraction, which was performed using HOTPANTS\footnote{\url{https://github.com/acbecker/hotpants}} \citep{1998ApJ...503..325A, 2015ascl.soft04004B}, with normalization to the science image and convolution to the template image. We present the host-subtracted $gri$-band photometry calibrated to the AB magnitude system in Figure \ref{fig:offset_lightcurve_all}.

Supplemental photometry in the $gr$-bands was obtained from the ZTF forced photometry server\footnote{\url{https://ztfweb.ipac.caltech.edu/cgi-bin/requestForcedPhotometry.cgi}} \citep{2019PASP..131a8003M}. We excluded data from both sets with magnitude errors greater than 0.1 mag. Las Cumbres subtractions with PS1 templates were compared with ZTF forced photometry to ensure similarity between subtraction methods, taking into account the expected variation between ZTF and Las Cumbres $r$-bands as measured in \cite{Newsome2023a}.

\subsection{Swift Photometry}\label{sec:swift}

UV observations were obtained with the Ultra\-violet Optical Telescope \citep[UVOT;][]{2005SSRv..120...95R} on board Swift in the \textit{UVW2}, \textit{UVM2}, \textit{UVW1}, $U$, $B$, and $V$ filters beginning MJD 59897.56, near the time of maximum light. A total of 14 epochs were observed through MJD 60103.0. The data were reduced with the Swift Ultraviolet/Optical Super\-nova Archive pipeline \citep{2014Ap&SS.354...89B}, using the aperture corrections and zero-points of \citet{2011AIPC.1358..373B}. Swift photometry is presented in Vega magnitudes alongside Las Cumbres and ZTF photometry in Figure \ref{fig:offset_lightcurve_all}. Because there are no archival images in the Swift UVOT bands of the host galaxy, these data are not host-subtracted.

Swift's X-ray Telescope (XRT) simultaneously observed AT 2022upj in photon-counting mode during the UVOT follow-up. Following the Swift XRT Data Reduction Guide\footnote{\url{https://www.swift.ac.uk/analysis/xrt/files/xrt_swguide_v1_2.pdf}}, we processed cleaned X-ray event files with \texttt{xselect}\footnote{\url{https://www.swift.ac.uk/analysis/xrt/xselect.php}} to create a light curve binned by observation where photons are selected within 40\arcsec\ of the event coordinates and the background is a region with no sources and a radius 150\arcsec. We use the online tool PIMMS\footnote{\url{https://cxc.harvard.edu/toolkit/pimms.jsp}} to convert these count rates into fluxes assuming a power law with a photon index of $\Gamma = 2.55$ at early times and $\Gamma = 3.1$ at late times (see \S \ref{sec:xrt_fits}). We display this light curve alongside the pseudo-bolometric luminosity of AT 2022upj (see \S \ref{sec:lightcurve}) as well as the evolving ratio between the two in Figure \ref{fig:offset_lightcurve_all}. 

\begin{figure}[t!]
    \includegraphics[scale=0.5]{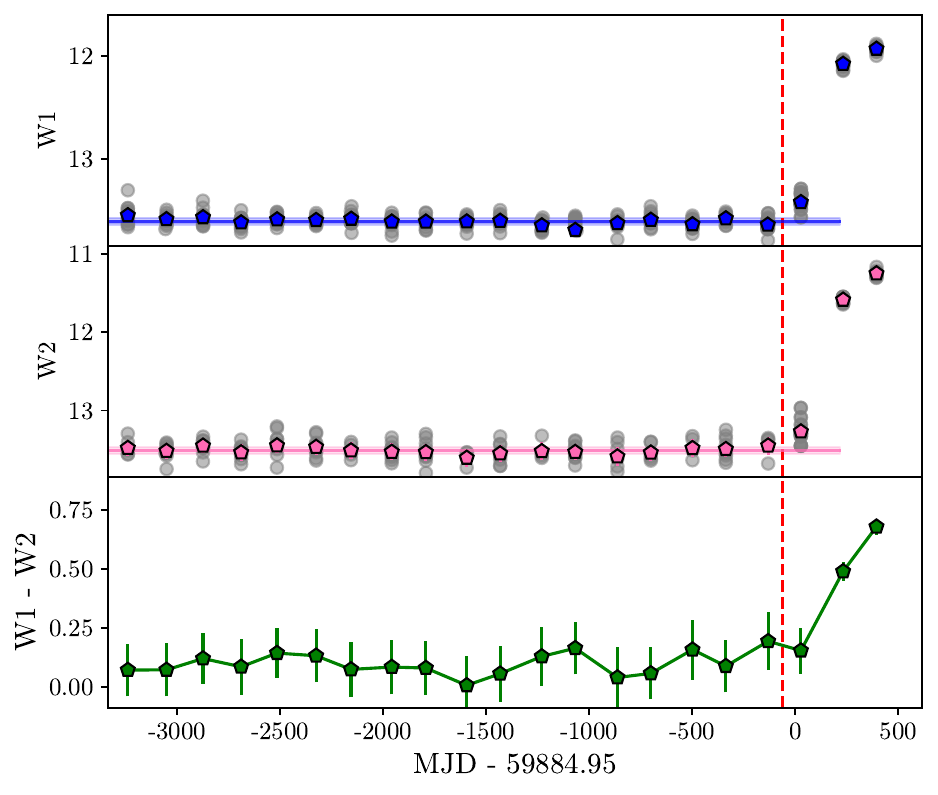}
    \caption{Unsubtracted WISE photometry at the position of AT 2022upj since 2014. Top: $W1$ $3.6\mu$m in blue circles and median quiescent value in a darker blue horizontal line; middle: $W2$ $4.5 \mu$m in pink circles and median quiescent value in a darker pink horizontal line; bottom: $W1 - W2$ colors in green pentagons. The host of AT 2022upj never approaches the expected AGN color-cut of $W1 - W2 > 0.8$ until the brightening in both bands after transient discovery at MJD 59822.39 (marked in the red vertical dashed line).}
    \label{fig:wise_lc}
\end{figure}

\subsection{WISE photometry}

The Wide-field Infrared Survey Explorer \citep[WISE;][]{2010AJ....140.1868W} first observed the host galaxy of AT 2022upj with the ALLWISE survey in 2010, and its observations resumed in 2013 when the survey reactivated as NEOWISE-R \citep[][]{Mainzer2011, 2014ApJ...792...30M}. We collected $W1$ ($3.4 \mu$m) and $W2$ ($4.6 \mu$m) data from the Infrared Sci\-ence Archive\footnote{\url{https://irsa.ipac.caltech.edu}}, producing 19 epochs of mid-infrared photo\-metry each separated by six months, one of which was taken after the discovery of AT 2022upj. We selected detections with good-quality frames ($\texttt{qi\_fact} > 1$), no contamination and confusion ($\texttt{cc\_flag} = 0$) and with magnitude errors $<0.15$ mag; filtering out $\sim 10\%$ of observations across all epochs. Figure \ref{fig:wise_lc} shows all epochs, including both raw data and weighted binned magnitudes per epoch, and the magnitude of the quiescent stage based on minimal variability prior to MJD 59822.39, when the transient was discovered. The quiescent magnitudes are measured from taking the median of all epochs per filter preceding the TDE detection. We also display the $W1 - W2$ color, which does not exceed the AGN color-cut of $W1 - W2 \geq 0.8$ \citep[][]{Stern2012} at any epoch.

\begin{figure*}[t!]
    \centering
    \includegraphics[scale=0.69]{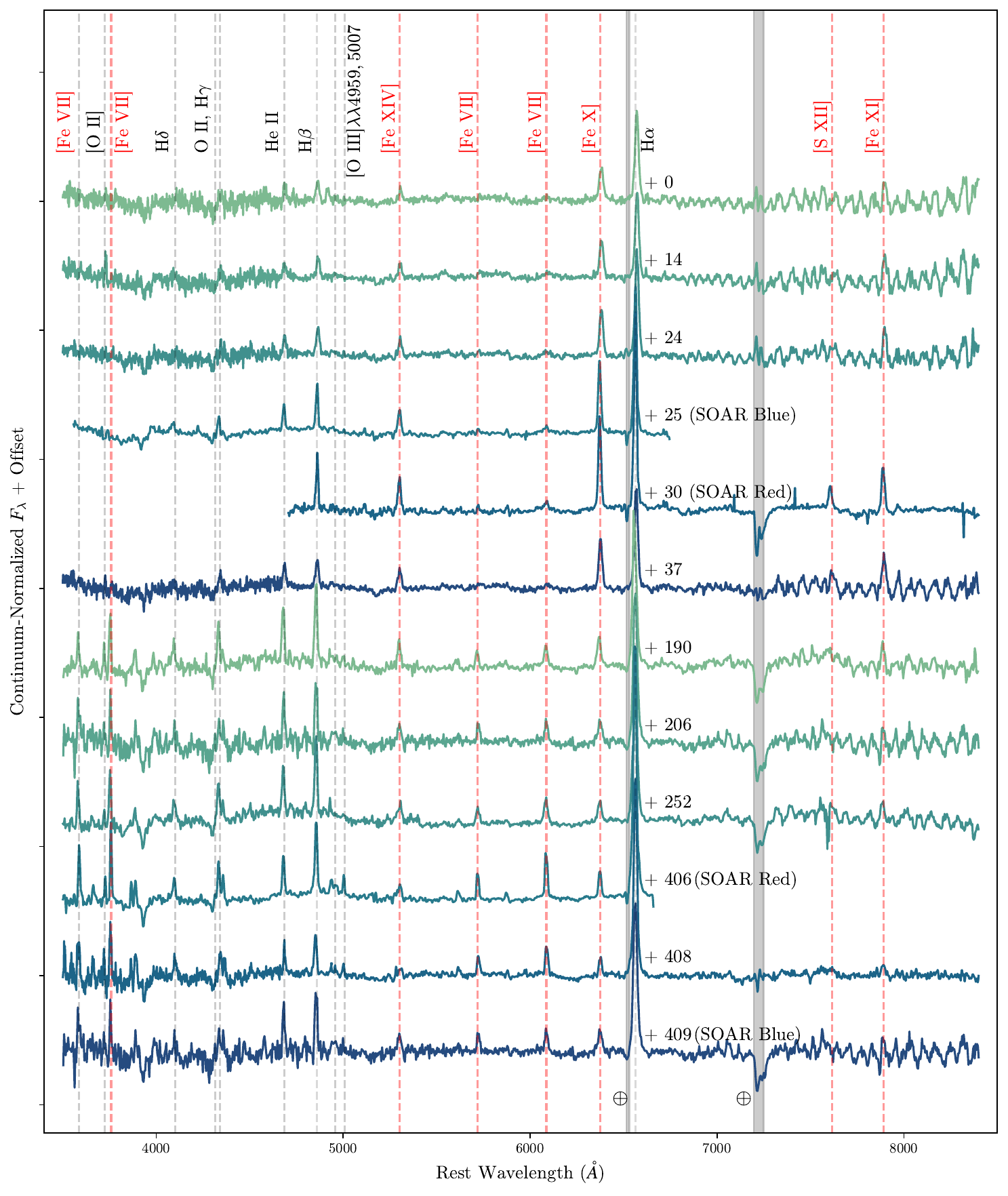}
    \caption{FLOYDS (unlabeled) and SOAR spectra of AT 2022upj, continuum-normalized, and offset with varying hues for clarity. Spectra are labeled by the number of rest-frame days from optical peak at which the observation was taken. We label the prominent narrow H$\alpha$, H$\beta$ and He II $\lambda$4686 lines in gray, and the high-ionization coronal lines ($[$\ion{Fe}{14}$]$ $\lambda$5303, $[$\ion{Fe}{7}$]$, $[$\ion{Fe}{10}$]$ $\lambda$6375, $[$\ion{S}{12}$]$ $\lambda$7611.0 and $[$\ion{Fe}{11}$]$ $\lambda$7891.8) in red. We also label the $[$\ion{O}{3}$]$ $\lambda$$\lambda$4959, 5007 doublet which is not evident in AT\,2022upj until after +190d from peak, when the $[$\ion{Fe}{7}$]$ also becomes prominent and the $[$\ion{Fe}{10}$]$ line weakens. The data behind this figure will be available on WISEReP.}
    \label{fig:spectra}
\end{figure*}

\subsection{Spectroscopy}\label{subsec:spec}

Eight optical spectra were taken with the FLOYDS instruments mounted on the Las Cumbres 2m Faulkes Telescopes North and South (FTN/FTS), using a 2\arcsec\ slit and reduced using the \texttt{floydsspec} pipeline\footnote{\url{https://github.com/svalenti/FLOYDS_pipeline/}}, as described in \cite{2014MNRAS.438L.101V}. The pipeline performs flux and wavelength calibration, cosmic-ray removal, and final spectrum extraction. The original spectra cover 3500 to 10,000 \AA, at a resolution $R \approx 300-600$ and epochs from $0$ to $+408$ days with respect to the $g$-band peak at MJD = 59884.95 (calculated by fitting a template TDE light curve to the observations with a least-squares method and finding the time of maximum light from the template; see \S \ref{sec:lightcurve}). Four additional spectra were obtained using the Goodman High Throughput Spectrograph on the Southern Astrophysical Research (SOAR) 4.1 m telescope, each using a 1\arcsec\ slit and the 400 mm$^{-1}$ line grating. Two used the blue camera in mode M1 (3000--7050 \AA), one used the red camera in mode M1 (3000--7050 \AA), and one used the red camera in mode M2 (5000-9050 \AA). We used the Goodman Spectroscopic Data Reduction Pipeline\footnote{\url{https://github.com/soar-telescope/goodman_pipeline}} to perform bias and flat correction and cosmic-ray removal, and then performed wavelength and sky-line calibration and extraction using PyRAF \citep{2012ascl.soft07011S}. We calibrated fluxes to a standard star observed on the same nights with the same instrumental setup. The dates and phases of each spectrum we report are listed in Table \ref{table:specdates}. We present all spectra and label the notable emission lines in Figure \ref{fig:spectra}.


\begin{table}[t]
\centering
\caption{The MJD and rest-frame phase at which each reported spectrum was taken by FLOYDS (Faulkes Telescope North, FTN, or Faulkes Telescope South, FTS) or SOAR, where phase is the number of rest-frame days from peak in $g$-band (at MJD 59884.95).}
\label{table:specdates}
\begin{tabular}{ccccc}
\hline
MJD & Phase (d) & Airmass & Exp. Time (s) & Telescope \\ 
\hline
59885 & +0 & 1.37 & 3600 & FTS\\
59900 & +14 & 1.62 & 3600 & FTS\\
59911 & +24 & 1.32 & 3600 & FTS\\
59912 & +25 & 1.23 & 1800 & SOAR\\
59917 & +30 & 1.22 & 1800 & SOAR\\
59925 & +37 & 1.55 & 3600 & FTS\\
60086 & +190 & 1.57 & 3600 & FTS\\
60103 & +206 & 1.62 & 3600 & FTS\\
60151 & +252 & 1.05 & 3600 & FTS\\
60313 & +406 & 1.49 & 2100 & SOAR\\
60316 & +408 & 1.49 & 3600 & FTN\\
60317 & +409 & 1.53 & 2100 & SOAR\\
\hline
\end{tabular}
\end{table}

\section{Results} \label{sec:results}

We assess the multi-band photometric data by comparing the light curve's late-time behavior against that of a ``typical" TDE, calculating blackbody radii and temperatures of the event across epochs, and using the light curve properties to determine the masses of the central BH and the stellar progenitor via assumptions of two different emission mechanisms.

\subsection{Slow Decline}\label{subsec:bump}

\begin{figure}[t!]
    \centering
    \includegraphics[scale=0.4]{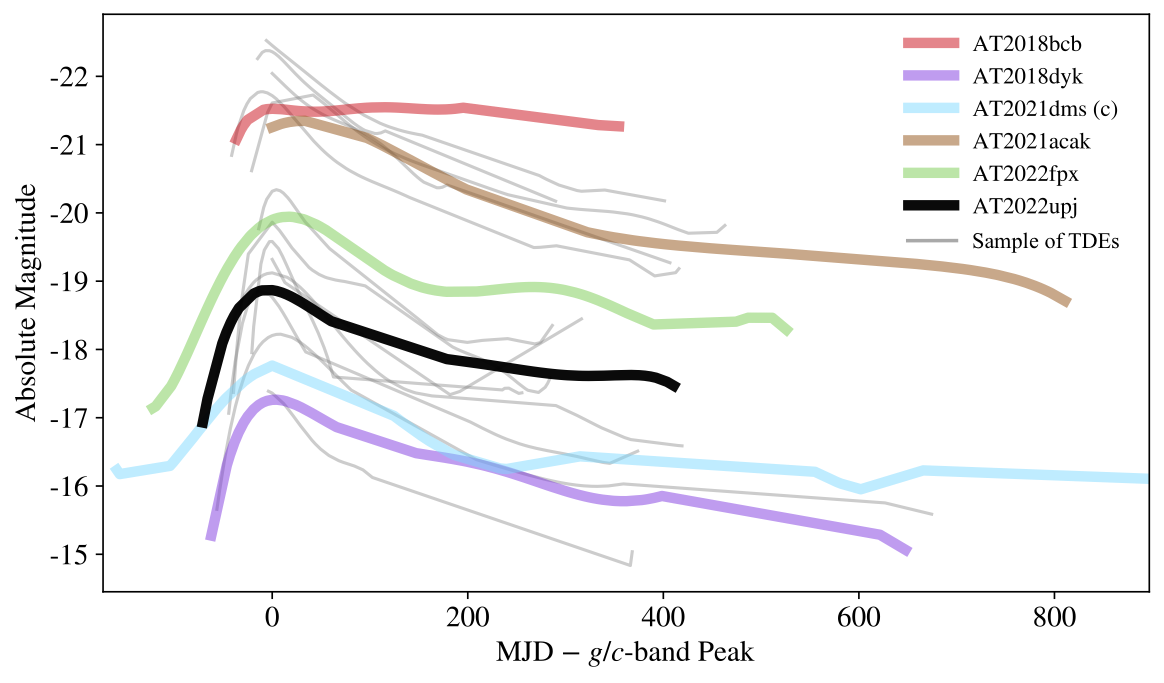}
    \caption{The $g$-band light curve of AT\,2022upj (black line) compared to that of a sample of TDEs (thin gray lines) for which data, with magnitude errors $<0.25$mag, exists beyond +200 days from the $g$-band peak \citep[AT\,2018hco, AT\,2018iih, AT2018\,jbv,  AT\,2018zr, AT\,2019ehz, AT\,2020mot, AT\,2020opy, AT\,2020qhs, AT\,2020ysg, AT\,2020yue, AT\,2021nwa, AT\,2022gri, AT\,2022lri, AT\,2022wtn; photometry all obtained from][] {VanVelzen21sample, Hammerstein2023, Yao2023, Neustadt2020, Hinkle2023, 2023RAA....23b5012L, Koljonen2024}. Other ECLEs concurrent with optical transients (AT\, 2018gn, AT\,2018bcb, AT\,2018dyk, AT\,2021dms, AT\,2021acak, AT\,2022fpx) are overplotted in color. AT\,2021dms is shown in the ATLAS $c$-band due to lack of ZTF coverage. All light curves are corrected for each object's redshift and shown in the rest-frame. The TDEs in the background sample show a median power law decline of $\alpha = 2.14 \pm 0.96$ prior to +200d from peak \citep[after which a plateau phase is common in TDEs;][]{Yao2023} while the ECLEs have a power law decline following $\alpha = 0.67 \pm 0.22$ over the same period, indicating a significant difference between the behavior of long-lasting TDEs versus ECLEs.}
    \label{fig:declinecomparison}
\end{figure}

We compare the $g$-band light curve of AT\,2022upj to a sample of TDEs and ECLEs for which there existed $g$-band (or ATLAS $c$-band) photometry after 200 days from each transient's peak \citep{VanVelzen21sample, Hammerstein2023, Yao2023, Neustadt2020, Hinkle2023, 2023RAA....23b5012L, Koljonen2024}, as shown in Figure \ref{fig:declinecomparison}. AT\,2022upj follows a power-law decline in the first 200d post $g$-band peak with $L \propto t^{-\alpha}$ where we find $\alpha=0.78 \pm 0.05$, substantially shallower than the typical TDE expectation of $\alpha = 5/3$. In fact, for all ECLEs with light curves detected over more than 250 days, we find an overall decline rate of $\alpha = 0.67 \pm 0.22$, while the sample of TDEs (that are not also ECLEs) collectively shows a decline with $\alpha = 2.14 \pm 0.96$ over the same period.

\subsection{Blackbody Fitting} \label{sec:lightcurve}

We fit the UV and optical spectral energy distribution (SED) of AT\,2022upj to a blackbody across its light curve to estimate the photospheric radius and temperature of the flaring region over time. We require five or more filters (ensuring only epochs with UV data) of data per epoch range (5 days) to fit a blackbody. Our fitting therefore starts from MJD = 59884.95 at $g$-band peak, using the \texttt{lightcurve\_fitting} package from \cite{griffin_hosseinzadeh_2022_6519623}.
\begin{deluxetable}{cccc}[t]\label{table:blackbody}
    \caption{MJD, bolometric luminosity, blackbody radius, and blackbody temperature for AT 2022upj determined by fitting a blackbody to epochs with Swift UVOT photometry alongside Las Cumbres and ZTF optical photometry.}
    \tablehead{
    \colhead{MJD} & \colhead{$L_\text{bol}$} & \colhead{$R$} & \colhead{$T$} 
    \\
    \colhead{} & \colhead{$10^{43}$ (erg s$^{-1}$)} & \colhead{$10^{-4}$ (pc)} & \colhead{(K)}
    }
    \startdata
        59898.9 & $ 5.83 \pm 0.29$ & $1.71 \pm 0.04 $& $ 23300 \pm 2500 $\\
        59910.5 & $ 5.61 \pm 0.33$ & $1.63 \pm 0.04 $& $ 23700 \pm 2500 $\\
        59913.5 & $ 6.18 \pm 0.41$ & $1.43 \pm 0.04 $& $ 25800 \pm 2500 $\\
        59939.2 & $ 4.25 \pm 0.20$ & $1.59 \pm 0.03 $& $ 22300 \pm 2500 $\\
        59947.2 & $ 4.12 \pm 0.50$ & $1.52 \pm 0.10 $& $ 22700 \pm 2500 $\\
        60091.8 & $ 1.48 \pm 0.08$ & $1.40 \pm 0.04 $& $ 18300 \pm 2500 $\\
        60248.9 & $ 1.00 \pm 0.09$ & $2.00 \pm 0.11 $& $ 13900 \pm 2500 $\\
    \enddata
\end{deluxetable}
\begin{figure}[b!]
    \centering
    \includegraphics[scale=0.475]{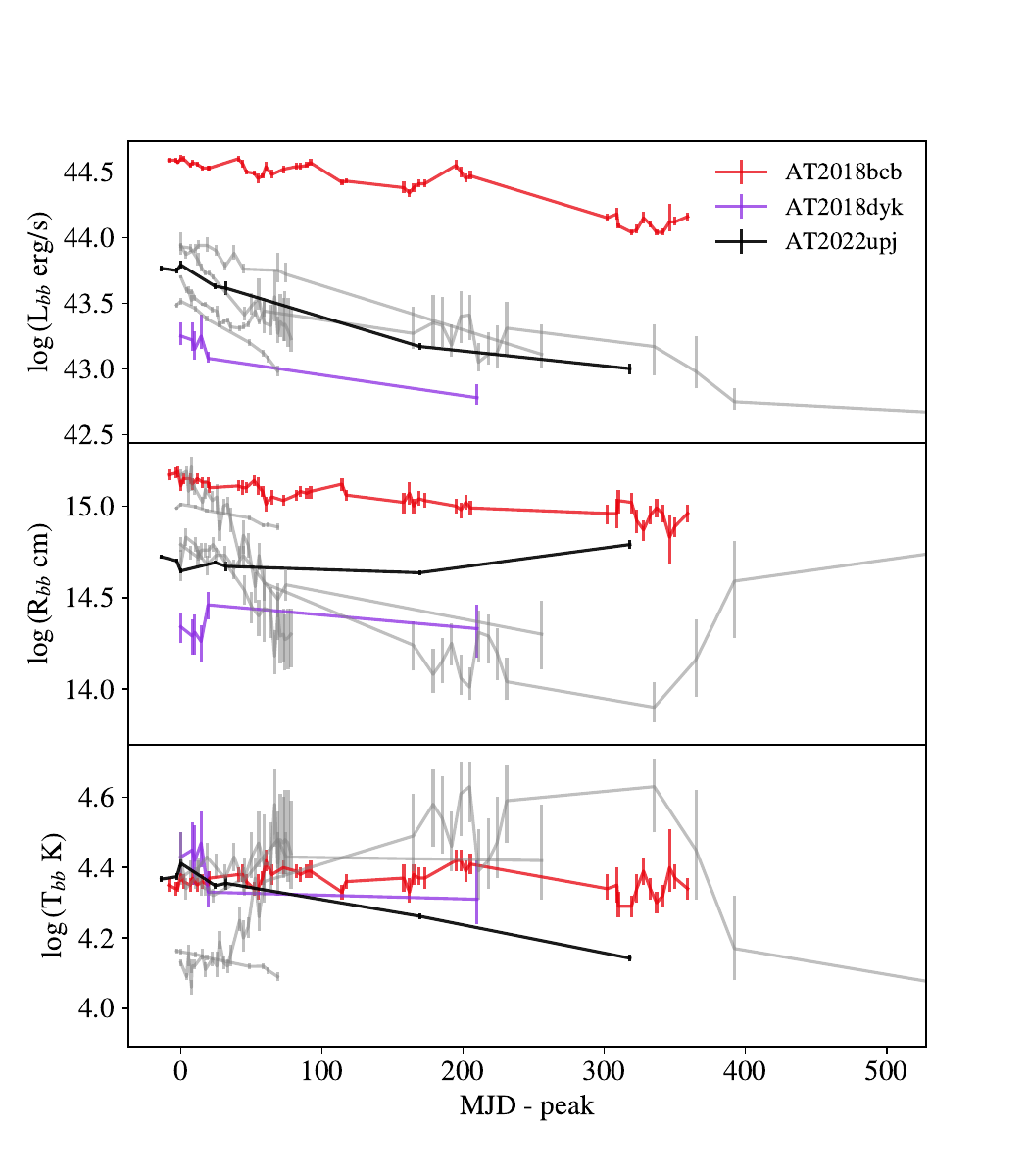}
    \caption{Blackbody physical properties inferred for AT\,2022upj (black lines) compared to TDEs (gray lines) and ECLEs (colored lines) for which blackbody data was publicly available (AT2018bcb, AT2018dyk, AT2018hco, AT2018zr, AT2019ehz, AT2020mot) Top: Blackbody luminosity in erg s$^{-1}$. Middle: Radius in cm. Bottom: Temperature in K. AT\,2022upj luminosity evolution is consistent with TDEs, and its peak radius and temperatures are also in line with TDEs, although there is deviation in both at late times.}
    \label{fig:blackbodyfit}
\end{figure}

The code uses Markov Chain Monte–Carlo (\texttt{MCMC}) sampling with the \texttt{emcee} package \citep{2013PASP..125..306F} with a broad log-flat prior of $10,000$ K $\leq T \leq 100,000$ K and $10 R_{\odot}$ $(\sim10^{-7} \text{pc}) \leq R \leq 10^{6} R_{\odot}$ $(\sim 10^{-2} \text{pc})$.

We use the host-subtracted photometry from Las Cumbres and ZTF, and non-subtracted Swift \textit{UVW1}, \textit{UVM2}, and \textit{UVW2} bands, in which host galaxy brightness is negligible (see \S \ref{sec:host_fits}).
\begin{figure*}[t!]
    \centering
    \includegraphics[scale=0.85]{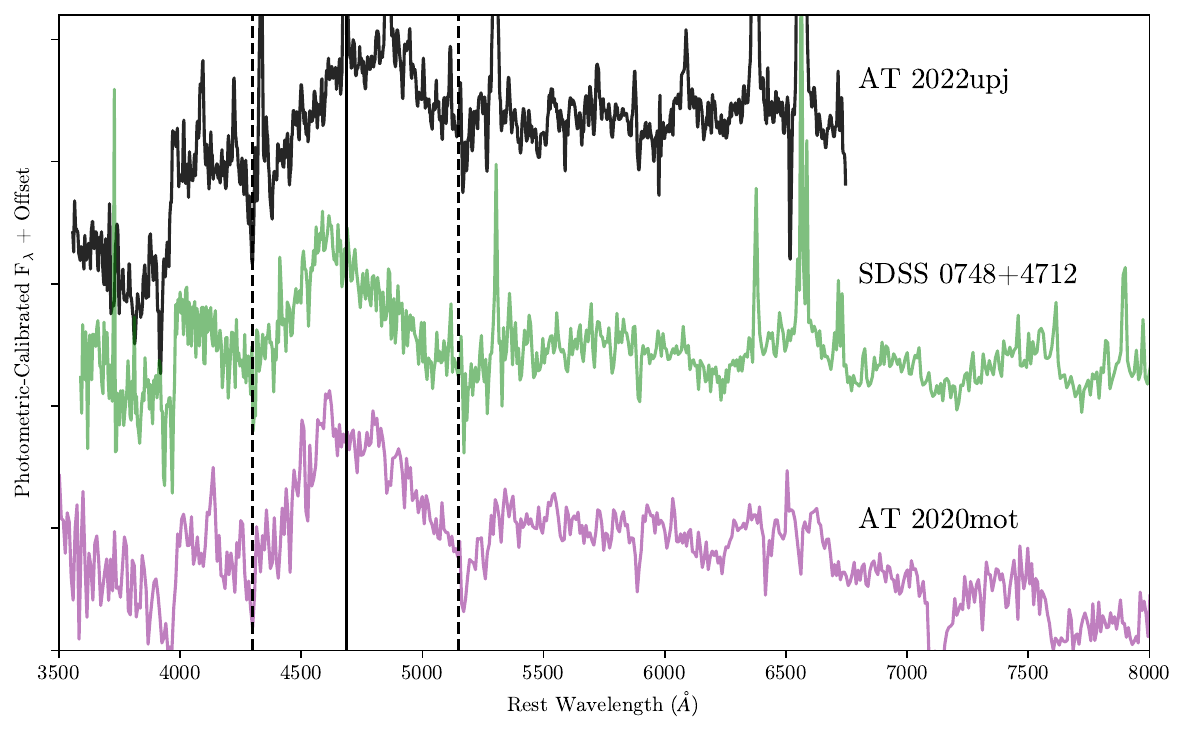}
    \caption{Close-up of the first SOAR spectrum of AT\,2022upj focusing on the broad He II $\lambda$4686 emission feature, with the same spectral segment shown for the extreme coronal line emitter SDSS 0748+4712 \citep[green;][]{Wang2011} and the TDE AT\,2020mot \citep[purple;][]{Newsome2023a}. SDSS 0748+4712 was the only ECLE from the SDSS sample which simultaneously showed broad He II, making AT\, 2022upj only the second instance of both features clearly detected, and the first for which the detections coincide with an observed optical transient.}
    \label{fig:HeII_zoom}
\end{figure*}
\begin{figure}[b!]
    \centering
    \includegraphics[scale=0.65]{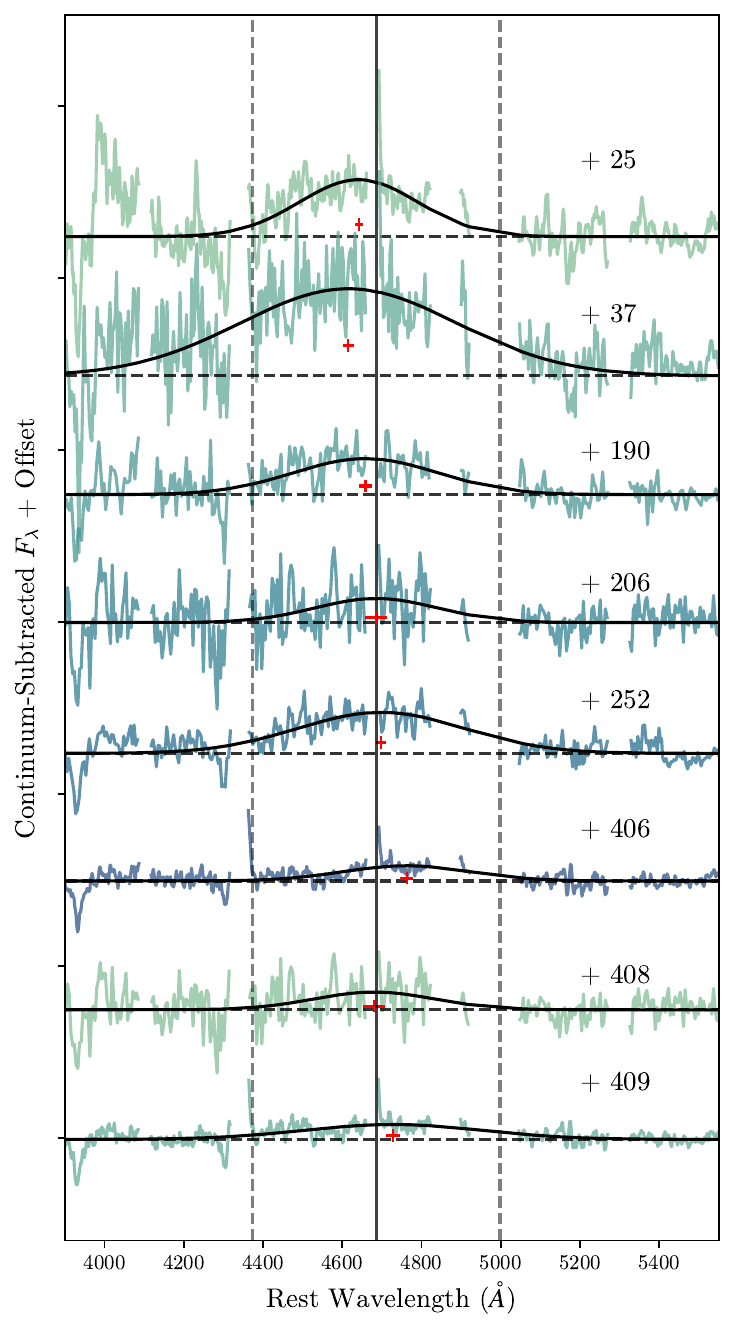}
    \caption{Spectral epochs of AT\,2022upj from +25d onward (rest-frame days since peak are noted next to each spectrum), each offset and differing in hue for clarity, in which the strong narrow emission lines have been masked to clearly show the remaining broad feature around He II $\lambda$4686, labeled at line center with a vertical black line and at $\pm$20,000 km s$^{-1}$ offset in vertical dashed lines. We also show the best-fit Gaussian to each line with the horizontal dashed line denoting the Gaussian zero level. We find the weighted average of the FWHM across all epochs is 14700 $\pm$ 3400 km s$^{-1}$.}
    \label{fig:HeII_all}
\end{figure}

Though this package provides errors on estimated parameters, we quantify our temperature measurements with the systematic uncertainties found in \cite{Arcavi2022} for the bounds on light curve fits to blackbodies with UV and optical data, as these values exceed the statistical errors found by the package. We find the temperature stays relatively constant between $T_{\text{BB}} = 24300\pm2500$ K and $T_{\text{\text{BB}}} = 25000\pm2500$ K over the first 60d.  The blackbody radius is found to decrease over the same period from $R_{\text{BB}} = 5.0\pm0.12\times 10^{14}$ cm to $R_{\text{BB}} = 3.8\pm0.12\times 10^{14}$ cm. After this early period the temperature decreases to a steady $T_{\text{BB}} \sim 15000$ K over 200d. These temperatures and radii are consistent with other TDE observations \citep[$T \sim 10000-50000$ K; e.g.][]{VanVelzen21}. We plot these values, and compare them to blackbody parameters that were publicly available for TDEs and ECLEs in our aforementioned sample, in Figure \ref{fig:blackbodyfit}.

\subsection{Black Hole and Pre-Disruption Stellar Masses} \label{subsec:mosfit}
\begin{deluxetable}{ccc}[t!]
\label{table:bh_masses}
    \caption{Estimates of the mass of the SMBH based on analyzing the UV-optical light curve as TDE emission with MOSFiT and TDEMass (which inherently assume different emission mechanisms) compared with the estimate from the scaling relation between black hole mass and bulge mass extended to the lower-mass regime by \cite{Scott2013}.}
    \tablehead{
    \colhead{Method} & \colhead{$M_\mathrm{BH}$} & \colhead{Source} \\
    \colhead{} & \colhead{($10^{6} M_{\odot}$)} & \colhead{}
    }
    \startdata
         MOSFiT & $0.61^{+0.1}_{-0.07}$  & \cite{Mockler2019}\\
         TDEMass & $1.6^{+0.48}_{-0.72}$ & \cite{Ryu2020}\\
         $M_\mathrm{BH}-M_\mathrm{bulge}$ & $0.83^{+1.56}_{-0.66}$ & \cite{Scott2013}\\
         $M_\mathrm{BH}-M_\mathrm{bulge}$ & $1.00^{+0.56}_{-0.36}$ & \cite{Greene2020}\\
    \enddata
\end{deluxetable}
Assuming the UV-optical flare is due to a TDE, we model the light curve using both the Modular Open Source Fitter for Transients, which attributes the emission to reprocessed accretion \citep[MOSFiT;][]{Guillochon2017, Mockler2019} and TDEMass \citep{Ryu2020}, which assumes the emission is from outer shocks of stream-stream collisions \citep{Piran2015}. The MOSFiT parameters are the black hole mass $M_\mathrm{BH}$, the mass of the star before disruption $M_{*}$, and eight other physical parameters detailed in \S \ref{sec:mosfit_outputs}, including the efficiency of the conversion of fallback mass to energy $\epsilon$ which can be used to account for additional emission sources. In contrast, TDEMass fits only the peak bolometric luminosity and corresponding temperature to analytically estimate the black hole and stellar masses. The detailed outputs of both fitting results are included in  \S \ref{sec:mosfit_outputs}.

We include photometry in all host-subtracted bands (i.e. all bands except $BV$, as for the blackbody fits) for fitting with MOSFiT's nested sampling method and find $M_\mathrm{BH} = 6.1^{+1.0}_{-0.7} \times 10^{5} M_{\odot}$ and a mass of $M_{*} = 0.32^{+0.44}_{-0.16} M_{\odot}$ for the disrupted star. In contrast, when using $L_\mathrm{bol} = 6.26\times 10^{43}$ erg s$^{-1}$ and {$T = 3.41\times 10^{4.2}$ K (interpolated values at the peak of the optical light curve), TDEMass finds $M_\mathrm{BH} = 1.6^{+0.48}_{-0.72} \times 10^{6} M_{\odot}$ and $M_{*} = 0.88^{+0.06}_{-0.07} M_{\odot}$. These values, while outside of one another's errors, indicate broadly a subsolar stellar progenitor and a SMBH that is likely less than $10^{7} M_{\odot}$, well within expectations and comparable to findings of other TDEs \citep[e.g.][]{Nicholl2022}.

As another measure of the SMBH mass, we use the \cite{Scott2013} relation between the stellar mass of Sérsic galaxies and their black hole mass by using the estimated stellar mass from the host SED fits (see \S \ref{sec:host_fits}). We choose this method due to the host's spheroidal nature, well fit with \texttt{GalSim}\footnote{\url{https://github.com/GalSim-developers/GalSim}} by a Sérsic profile with index $n > 6.0$ indicating strong central concentration of brightness. As a check on this conclusion we also fit the host image with \texttt{photutils} and find an ellipticity of 0 fits the host with minimal residuals. Therefore, we proceed with using the stellar mass as the bulge mass and find $M_\mathrm{BH} = 0.83^{+1.56}_{-0.66} \times 10^{6} M_{\odot}$, in agreement with the former two estimates which assume a TDE nature of the UV-optical light curve. As a further check we also use the same values in the $M_\mathrm{BH}-M_\mathrm{bulge}$ relation from \cite{Greene2020} and find $M_\mathrm{BH} = 1.00^{+0.56}_{-0.36} \times 10^{6} M_{\odot}$ again agreeing with all prior estimates.

\subsection{X-ray Properties}

The X-ray light curve of AT 2022upj reveals a relatively constant flux between optical peak and +200d later, when Swift XRT data is binned per observation and requiring a minimum of 4 counts per epoch, as shown in Figures \ref{fig:offset_lightcurve_all} and \ref{fig:coronal_lightcurve}. The data prior to +200d are either $3\sigma$ detections or upper limits, so it is possible that the X-ray emission was variable around the detection limit over this period.

The X-ray flux over time does not exceed $2\%$ of the bolometric luminosity until after +300d from optical peak. This is similar to the extreme coronal line emitter AT 2018bcb \citep[ASASSN-18jd;][]{Neustadt2020}. However, our data does not indicate any significant softening nor hardening with time (see below), and the luminosity evolution shows no sudden spikes in brightness as all epochs are within one another's errors, so we can only infer generally constant evolution through +300d from optical peak. The observed X-ray luminosity stays within $L_{\text{XRT}} \sim 10^{41} - 10^{42}$ erg s$^{-1}$ which is comparable to X-ray selected TDEs \citep{Gezari2021}. Afterwards there is a clear increase in X-ray luminosity at +350d, similar to what was seen for the TDE ASASSN-15oi \citep{Gezari2017}, which is maintained over all subsequent observations. The reason for this evolution may be the onset of accretion, an accretion state change, or the emergence of pre-existing accretion that had been previously obscured, but analysis of the late X-ray emission will be presented in future work.

We bin the X-ray data into two epochs labeled ``pre-burst" and ``burst'', to measure the X-ray spectrum at these two stages, and fit each with power law and blackbody models using \texttt{XSPEC} and the corresponding online data products tool \citep[][see fits in \S \ref{sec:xrt_fits}]{2009MNRAS.397.1177E}. For this spectroscopic fitting we include all counts detected within each bin. The spectra are better fit by the power-law model which has substantially lower W-stat values for each epoch. The power law fits find an increasing photon index, from $\Gamma = 2.55^{+1.37}_{-0.35}$ near peak to $\Gamma = 3.08^{+0.23}_{-0.21}$ at late times, consistent with no change within the errors. Via blackbody fitting we find an X-ray temperature estimate of $200 \pm 50$ eV across both peak and later times, which is within the range observed in X-ray selected TDEs \citep{Gezari2021}, but we note that the data is better fit by the aforementioned power laws. The weak early-time X-ray detections are direct evidence for the presence of a simultaneous ionizing source during the optical flare energetic enough to produce the forbidden iron lines.

\begin{figure*}[t!]
    \centering
    \includegraphics[scale=0.65]{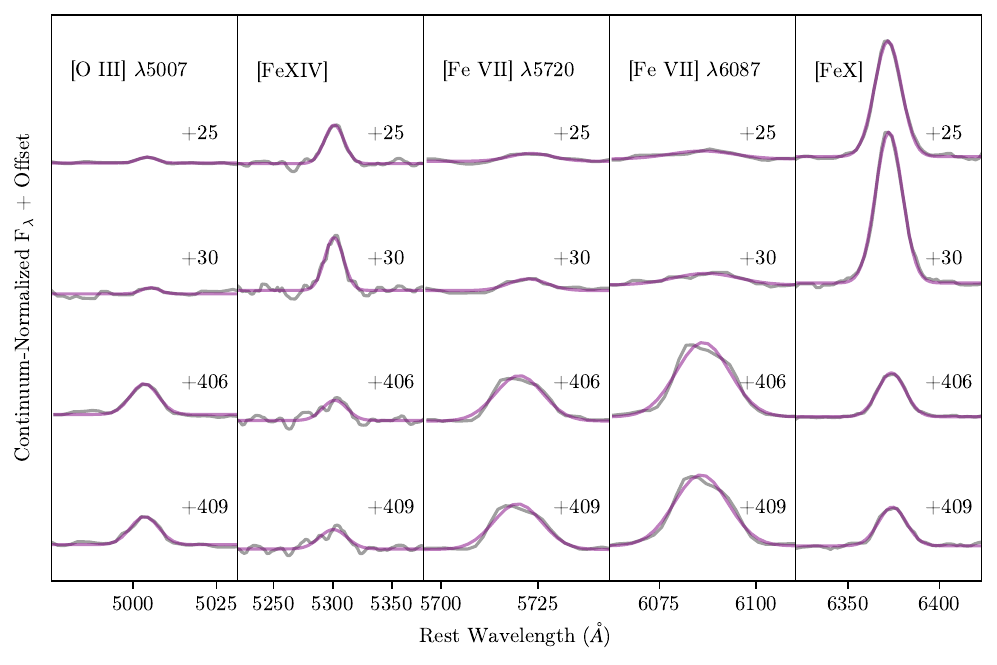}
    \includegraphics[scale=0.65]{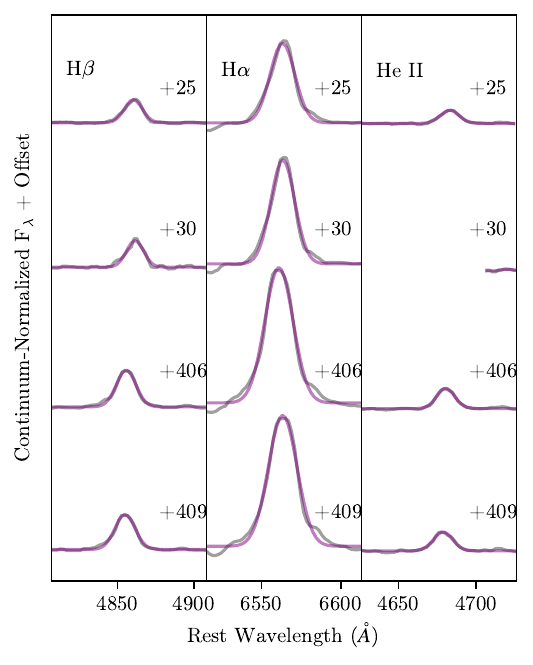}
    \caption{Gaussian fits for key emission lines across the four higher-resolution SOAR spectra taken at +25, +30, +406, and +409 rest-frame days from optical peak. Left: High-ionization lines $[$\ion{O}{3}$]$ $\lambda$5007, $[$\ion{Fe}{14}$]$ $\lambda$5303, $[$\ion{Fe}{7}$]$ $\lambda$5720, $[$\ion{Fe}{7}$]$ $\lambda$6087, and $[$\ion{Fe}{10}$]$ $\lambda$6375. Right: H$\beta$, H$\alpha$, and the narrow component of He II, for which we treat the broad component as a local continuum. The gray lines are the photometrically-calibrated data offset from one another for clarity, and the overlaying purple lines are the Gaussian fits to the data. The first two detections of $[$\ion{O}{3}$]$ and $[$\ion{Fe}{7}$]$ are less than 3-$\sigma$ detections, as are the last two observations of $[$\ion{Fe}{14}$]$.}
    \label{fig:gaussianfits}
\end{figure*}

Archival GALEX measurements used in host galaxy SED fitting (see \ref{sec:host_fits}) indicate no X-ray detection, down to 1.92 $\times 10^{-31}$ erg s$^{-1}$cm$^{-2}$, 10 years prior to the transient. Therefore we proceed with the assumption that the detected X-ray emission is caused by a TDE as opposed to background host emission from an AGN. However we note that, since our earliest observations show coronal lines without first showing quiescence, we cannot exclude the possibility that the host became active over the last ten years.

\begin{deluxetable}{cccc}[t!]
\label{table:virialdata}
    \caption{Central wavelength, FWHM, and integrated flux of the broad He II $\lambda$4686 for each epoch in which it is observed (+25d onward), as found via Gaussian fitting.}
    \tablehead{
    \colhead{Epoch} & \colhead{Wavelength} & \colhead{FWHM} & \colhead{Flux} \\
    \colhead{} & \colhead{(\AA)} & \colhead{(km s$^{-1}$)} & \colhead{($10^{-15}$ erg s$^{-1}$ cm$^{-2}$)}
    }
    \startdata
        +25 & 4641 $\pm$ 10 & 11400 $\pm$ 800 & 12.4 $\pm$ 0.17\\
        +37 & 4615 $\pm$ 13 & 20900 $\pm$ 1000 & 34.7 $\pm$ 0.27\\
        +190 & 4658 $\pm$ 15 & 13800 $\pm$ 1200 & 9.58 $\pm$ 0.19\\
        +206 & 4685 $\pm$ 28 & 11600 $\pm$ 2100 & 5.42 $\pm$ 0.48\\
        +252 & 4698 $\pm$ 12 & 15900 $\pm$ 900 & 12.6 $\pm$ 0.13\\
        +406 & 4762 $\pm$ 16 & 11900 $\pm$ 1200 & 3.58 $\pm$ 0.10 \\
        +408 & 4681 $\pm$ 27 & 11600 $\pm$ 2100 & 3.97 $\pm$ 0.34\\
        +409 & 4728 $\pm$ 17 & 16900 $\pm$ 1300 & 4.90 $\pm$ 0.09\\
    \enddata
\end{deluxetable}

\subsection{Spectral Evolution}\label{sec:linefits}

The spectra of AT 2022upj indicate the nuclear transient flare's interaction with circumnuclear material, firmly putting the transient within the class of extreme coronal line emitters. Importantly, the spectra also show broad He II $\lambda$4686 as shown in Figures \ref{fig:HeII_zoom} and \ref{fig:HeII_all}, a clear TDE signature (as defined by \citealt{Gezari2012} and \citealt{Arcavi2014}). We fit the broad He II component at each epoch using a Gaussian with free paramaters for the continuum, amplitude, and central wavelength. We measure a He II velocity of $14700 \pm 3400$ km s$^{-1}$ using a weighted average of the FWHM from the Gaussian fits to each epoch in which He II is detected (from +25d onward) as shown in Figure \ref{fig:HeII_all}\footnote{The He II line may be blended with \ion{N}{3} $\lambda\lambda 4100,4640$, as identified in some TDEs \citep[e.g.][]{Leloudas2019}, which would not be resolved by our observations.}. This velocity is in agreement with the FWHMs of He II in TDEs found by \cite{Charalampopoulos2022}. The slightly blueshifted central wavelength of the Gaussian fit to the epochs nearest optical peak (+25d and +37d) are also in agreement with TDE profiles \citep{Arcavi2014, Charalampopoulos2022}. The strength of the broad He II declines slowly over time but remains evident above the continuum in each epoch. This is the first event to show both broad He II and highly-ionized coronal lines simultaneously since SDSS J0748+4712 \citep{Wang2011}, here further coinciding with the peak of a UV-optical flare. AT 2022upj is thus the first optical transient to strongly link ECLEs with TDE origins.

The first spectrum of AT 2022upj, taken at optical peak, shows strong narrow $[$\ion{Fe}{10}$]$ $\lambda$6375 and $[$\ion{Fe}{14}$]$ $\lambda$5303 lines. The initial absence of [O III] emission differentiates AT 2022upj from AGN, as coronal line strengths measured by integrated flux in Seyfert galaxies are normally only a few percent of the [O III] doublet line strength \citep{Nagao2000}.

The first three spectra taken by FLOYDS show little  evolution in line presence and strength. Afterwards the higher-resolution SOAR blue and red spectra, taken 5 days apart at +25 and +30 rest-frame days from peak respectively, show weak $[$\ion{Fe}{7}$]$ $\lambda$5720 and $\lambda6087$ in addition to the strong $[$\ion{Fe}{10}$]$ $\lambda$6375 and $[$\ion{Fe}{14}$]$ $\lambda$5303 lines. The SOAR spectra also confirm the presence of [S XII] $\lambda7611$ and [Fe XI] $\lambda7892$ by +30d after peak. The red fringing in the FLOYDS spectra unfortunately make the [S XII] lines indistinguishable from the continuum, however the [Fe XI] line is strong enough in each epoch to observe clearly despite fringing, but not to resolve its equivalent width.

We measure the photometrically calibrated fluxes and equivalent widths of the high-ionization lines that are observed in every spectrum: $[$\ion{Fe}{14}$]$ $\lambda$5303, $[$\ion{Fe}{7}$]$ $\lambda$5720 (labeled ``[Fe VIIa]" in figures), $[$\ion{Fe}{7}$]$ $\lambda$6087 (``[Fe VIIb] in figures), and $[$\ion{Fe}{10}$]$ $\lambda$6375. We fit a univariate spline to the continuum around each line, normalize to the continuum level, and fit a Gaussian to each line at each epoch to measure the FWHM and fluxes. We show the Gaussian fits to the narrow emission lines from the higher-resolution SOAR spectra in Figure \ref{fig:gaussianfits}. We note that the first two detections of $[$\ion{O}{3}$]$ and $[$\ion{Fe}{7}$]$ are less than 3-$\sigma$ detections, as are the last two observations of $[$\ion{Fe}{14}$]$, thus we do not infer FWHM values from these measurements.

Comparing the steady H$\alpha$ flux to that of each high-ionization line, shown in Figure \ref{fig:coronal_lightcurve}, reveals intriguing trends. Specifically, $[$\ion{Fe}{10}$]$ $\lambda$6375 and $[$\ion{Fe}{14}$]$ $\lambda$5303 weaken after +190 days while $[$\ion{Fe}{7}$]$ $\lambda$5720 and $\lambda6088$ are only first detected above a 3$-\sigma$ level at this point, and strengthen afterwards.

Finally, we emphasize that $[$\ion{O}{3}$]$ $\lambda\lambda$4959,5007 clearly emerges only after +400d from peak, when the coronal lines are dominated by $[$\ion{Fe}{7}$]$ while $[$\ion{Fe}{10}$]$ has weakened. \cite{Wang2011, Wang2012} also observed an increase in [O III] flux after the fading of high-ionization coronal lines in their ECLE samples. There, the change took place over many years, while the change in AT\,2022upj is visible within 13 months.

\begin{figure}[t!]
    \includegraphics[scale=0.8]{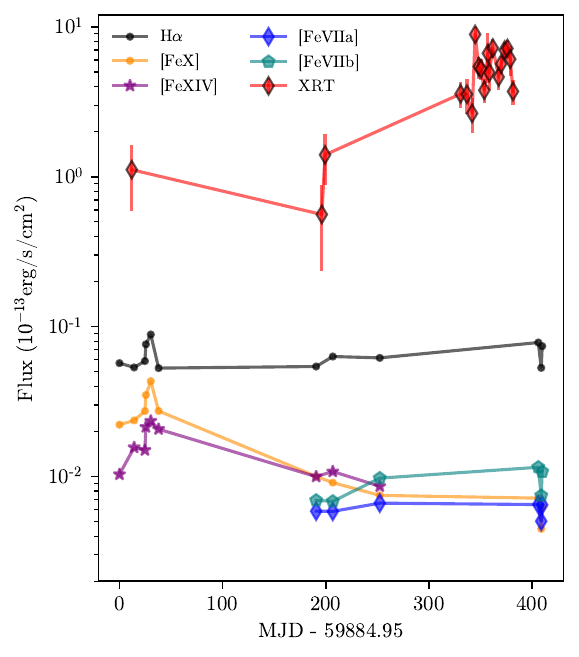}
    \caption{X-ray flux (red diamonds) from Swift XRT alongside the photometrically-calibrated fluxes of the prominent emission lines $[$\ion{Fe}{10}$]$ $\lambda$6375 (dark orange small dots), $[$\ion{Fe}{14}$]$ $\lambda$5303 $\lambda5303$ (purple stars), ``[Fe VIIa]" $\lambda$5720 (blue diamonds), ``[Fe VIIb]" $\lambda6087$ (green pentagons), and H$\alpha$ $\lambda$6563 (black dots).}
    \label{fig:coronal_lightcurve}
\end{figure}

\section{Discussion} \label{sec:gas}

\begin{figure}[t!]
    \includegraphics[scale=0.8]{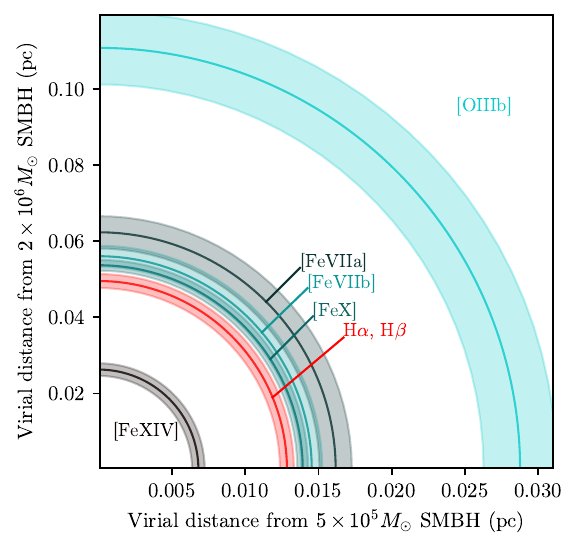}
    \caption{Map of the distance between the coronal-line-emitting circumnuclear gas and the SMBH based on the line FWHM and the assumption of virialized motion. Using only the FWHMs measured from Gaussian fits of the coronal lines in the higher-resolution SOAR spectra, the gas shows consistent velocities across all epochs. Within the range of black hole masses found for AT\,2022upj by light curve fitting and $M_\mathrm{BH}-M_\mathrm{bulge}$ relations, the gas is situated within 0.1pc of the SMBH.}
    \label{fig:virialdistances}
\end{figure}

The ionization potentials of 99 eV for $[$\ion{Fe}{7}$]$ \citep{Kramida_2022}, 235 eV for $[$\ion{Fe}{10}$]$ $\lambda$6375, 262 eV for [Fe XI], and 361 eV for $[$\ion{Fe}{14}$]$ $\lambda$5303 \citep{1987ApJ...318..145F} necessitate a strong ionizing continuum in the regime of extreme ultraviolet to soft X-rays. Our XRT observations show a weak X-ray flare (Figure \ref{fig:offset_lightcurve_all}) and an overall soft spectrum ($\Gamma \sim 3.0$) over the first 300d from optical peak, such that the fraction of flux above the necessary ionizing potentials as measured from the power-law fit is 40--48$\%$ of the total measured X-ray flux. This fraction of X-ray flux is still 22.3--26.3 times brighter than the measured strength of the coronal lines seen 14.5d later, providing more than enough of the ionizing photons to explain the line fluxes.

\begin{deluxetable}{cccc}[b]\label{table:virialdata}
    \caption{Median FWHM for each high-ionization coronal line, $[$\ion{O}{3}$]$ $\lambda$5007, and combined H$\alpha$ and H$\beta$ across SOAR-measured spectral epochs, and the corresponding virial distance for each line, where $R_\mathrm{min}$ corresponds to the distance assuming $M_\mathrm{BH} = 5\times10^{5} M_{\odot}$ and $R_\mathrm{max}$ corresponds to the distance assuming $M_\mathrm{BH} = 2\times10^{6} M_{\odot}$. These distances correspond to light travel times of 10--80 days.}
    \tablehead{
    \colhead{Line} & \colhead{FWHM} & \colhead{$R_\mathrm{min}$} & \colhead{$R_\mathrm{max}$} \\[-6pt]
    \colhead{} & \colhead{(km s$^{-1}$)} &\colhead{(10$^{16}$cm)} &\colhead{(10$^{16}$cm)}
    }
    \startdata
        $[$\ion{Fe}{14}$]$ & 556.4 $\pm$ 20.8 & 2.32 $\pm$ 0.2 & 8.92 $\pm$ 0.6\\
        $[$\ion{Fe}{7}a$]$ & 371.0 $\pm$ 12.4 & 5.21 $\pm$ 0.3 & 20.06 $\pm$ 1.3\\
        $[$\ion{Fe}{7}b$]$\ & 400.8 $\pm$ 11.0 & 4.46 $\pm$ 0.2 & 17.19 $\pm$ 0.9\\
        $[$\ion{Fe}{10}$]$ & 406.6 $\pm$ 4.1 & 4.33 $\pm$ 0.1 & 16.69 $\pm$ 0.3\\
        $[$\ion{O}{3}b$]$ & 278.9 $\pm$ 23.9 & 9.21 $\pm$ 1.4 & 35.48 $\pm$ 5.4\\
        H$\alpha$, H$\beta$ & 439.9 $\pm$ 8.8 & 3.70 $\pm$ 0.1 & 14.26 $\pm$ 0.6\\
    \enddata
\end{deluxetable}

In \S \ref{subsec:mosfit} we find BH mass estimates of $M_\mathrm{BH} \sim 0.5-2\times 10^{6} M_{\odot}$. Using this approximate range, we can determine lower and upper distance estimates of the coronal line-emitting gas. We use the FWHM from the Gaussian fits of each line measured in the SOAR spectra\footnote{We do not use the FLOYDS spectra here as their resolution is $\sim$500 km s$^{-1}$, whereas the SOAR resolution is $\sim$150 km s$^{-1}$.} as a measure of the virial velocity of the emitting gas and thus find the virial radius for each line. The median FWHM for each line across all epochs, and each line's corresponding virial distances, are listed in Table \ref{table:virialdata}. The velocity found for $[$\ion{Fe}{10}$]$ is faster than the same line FWHM measured in AT\,2019qiz \citep[273--368 km s$^{-1}$;][]{Short2023} and VT\,1008 \citep[212--370 km s$^{-1}$;][]{Somalwar2023}.

The inferred radii correspond to light crossing times of 10--80 days, consistent within the first detection by ZTF which was 57 rest-frame days prior to peak, given that X-ray activity prior to the first Swift detection near peak cannot be ruled out.

Figure \ref{fig:virialdistances} shows a schematic of where the emitting material is located relative to the SMBH based on the virial distance estimates for the two ends of the black hole mass estimate. In either case we see the closest material is within 0.1pc of the SMBH. We note that we cannot use reverberation mapping between the X-ray and the emergence of coronal lines to also infer the light-crossing time, as the first XRT observation was made +11 rest-frame days after coronal lines were already observed. The initial X-ray flare we observe may have peaked prior to observations, and if so, the coronal lines seen at optical peak may be indicative of an earlier X-ray flare that peaked before the optical bands \citep{Komossa2008}.

We add that the spectroscopic changes of AT\,2022upj occur while the WISE fluxes (as shown in Figure \ref{fig:wise_lc}) were still increasing as a likely dust echo of the initial flare, indicating a stratified circumnuclear environment. Taking the most recent NEOWISE observation at +466d from optical peak to be a lower limit on the peak time of the dust echo \citep[and therefore as the reverberation lag time;][]{Jiang2016, VanVelzen2016, Jiang2021}, the dust is inferred at a distance of at least 0.40pc from the SMBH. Thus AT\,2022upj is the first case of an ECLE-TDE whose multilayered circumnuclear material has been unveiled by the combination of multiband photometry and short-term spectroscopic followup.

\section{Conclusions}\label{sec:conclusion}

We present AT 2022upj as the first confirmed ECLE-TDE with concurrent X-ray emission, an optical-UV flare and evolving emission lines over $\sim$year-long timescales. The association of ECLEs with TDEs has been assumed due to the energetics required which rule out supernovae and AGN, but the possibility of AGN accretion state changes and the growing sample of ambiguous nuclear transients made this connection tenuous, until now. The detection of broad He II $\lambda$4686 here provides evidence for a TDE origin of ECLE signatures.

Our XRT observations provide the earliest detection of X-ray emission in an ECLE during an optical transient, providing more than enough ionizing flux to explain the extreme coronal lines. Furthermore, the virial motion of the coronal lines around the central SMBH of mass $M_\mathrm{BH} \sim 0.5-2\times 10^{6} M_{\odot}$ reveals that the closest emitting material is situated within 0.1pc of the SMBH. These findings, in conjunction with the dust echo detected in WISE mid-IR bands indicating dust at a minimum of 0.4pc, confirm the detection of stratified circumnuclear material around a SMBH at among the smallest scales yet probed. Finally, the emergence of $[$\ion{O}{3}$]$ alongside the increase in X-ray brightness may indicate a dramatic change in accretion state or the unveiling of pre-existing accretion from obscuring material, providing an opportunity to investigate the emission mechanisms of TDEs\citep[e.g.][]{Dai2018, Ryu2020}.

Our observations and analyses of AT 2022upj unveil the smallest-scale circumnuclear gas around a SMBH while highlighting the intricacies of emission from a tidal disruption event in a gas-rich and dusty environment. We have shown the power of early-time spectroscopic monitoring of nuclear transients to probe extreme and obscured extragalactic environments. We encourage earlier monitoring of TDEs for signs of coronal line emission and immediate follow-up with X-ray telescopes to identify the ionizing source and constrain the emission mechanism of the central flare.

\vspace{1cm}
The Las Cumbres Observatory group is supported by NSF grants AST-1911151 and AST-1911225. I.A., Y.D. and S.F. acknowledge support from the European Research Council (ERC) under the European Union Horizon 2020 research and innovation program (grant No. 852097). I.A. further acknowledges support from the Israel Science Foundation (grant No. 2752/19),from the United States—Israel Binational Science Foundation (BSF; grant number 2018166), and from the Pazy foundation (grant number 216312). This research makes use of observations from the Las Cumbres Observatory global telescope network as well as the NASA/IPAC Extragalactic Database (NED), which is operated by the Jet Propulsion Laboratory, California Institute of Technology, under contract with NASA.

%

\vspace{5mm}
\facilities{LCO, ZTF, SOAR, Swift (XRT and UVOT), WISE}


\software{astropy \citep{2013A&A...558A..33A,2018AJ....156..123A}, lcogtsnpipe \citep{Valenti16}, Light Curve Fitting \citep{griffin_hosseinzadeh_2022_6519623}, PyRAF \citep{2012ascl.soft07011S}, MOSFiT \citep{Guillochon2017}, TDEMass \citep{Ryu2020}, HOTPANTS \citep{1998ApJ...503..325A, 2015ascl.soft04004B}, HEAsoft \citep{2014ascl.soft08004N}, \texttt{BAGPIPES} \citep{Carnall2018}}

\bibliography{main}{}
\bibliographystyle{aasjournal}
 
\appendix

\section{XRT Power-Law Fits}
\label{sec:xrt_fits}

We use HEASOFT v6.32\footnote{\url{https://www.swift.ac.uk/user_objects/}} \citep{2014ascl.soft08004N} to extract source and background count rates for each observation using a source region with a radius of 40\arcsec\ centered and background region with radius of 150\arcsec. All count rates were corrected for encircled energy fraction. The results of the fits are in Figure \ref{fig:xrt_powerlaw}.

\begin{figure}
    \centering
    \includegraphics[scale=0.3]{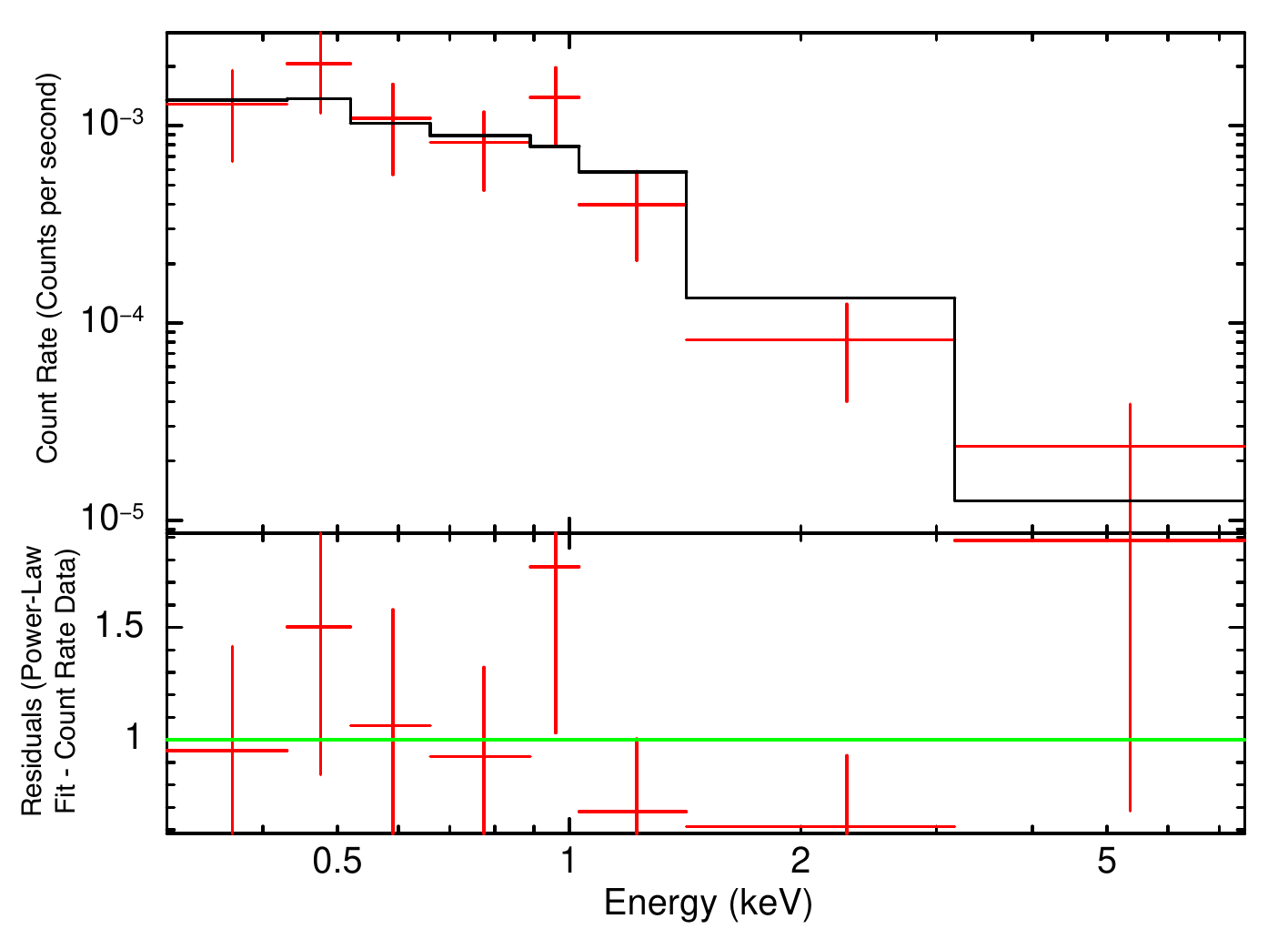}
    \includegraphics[scale=0.3]{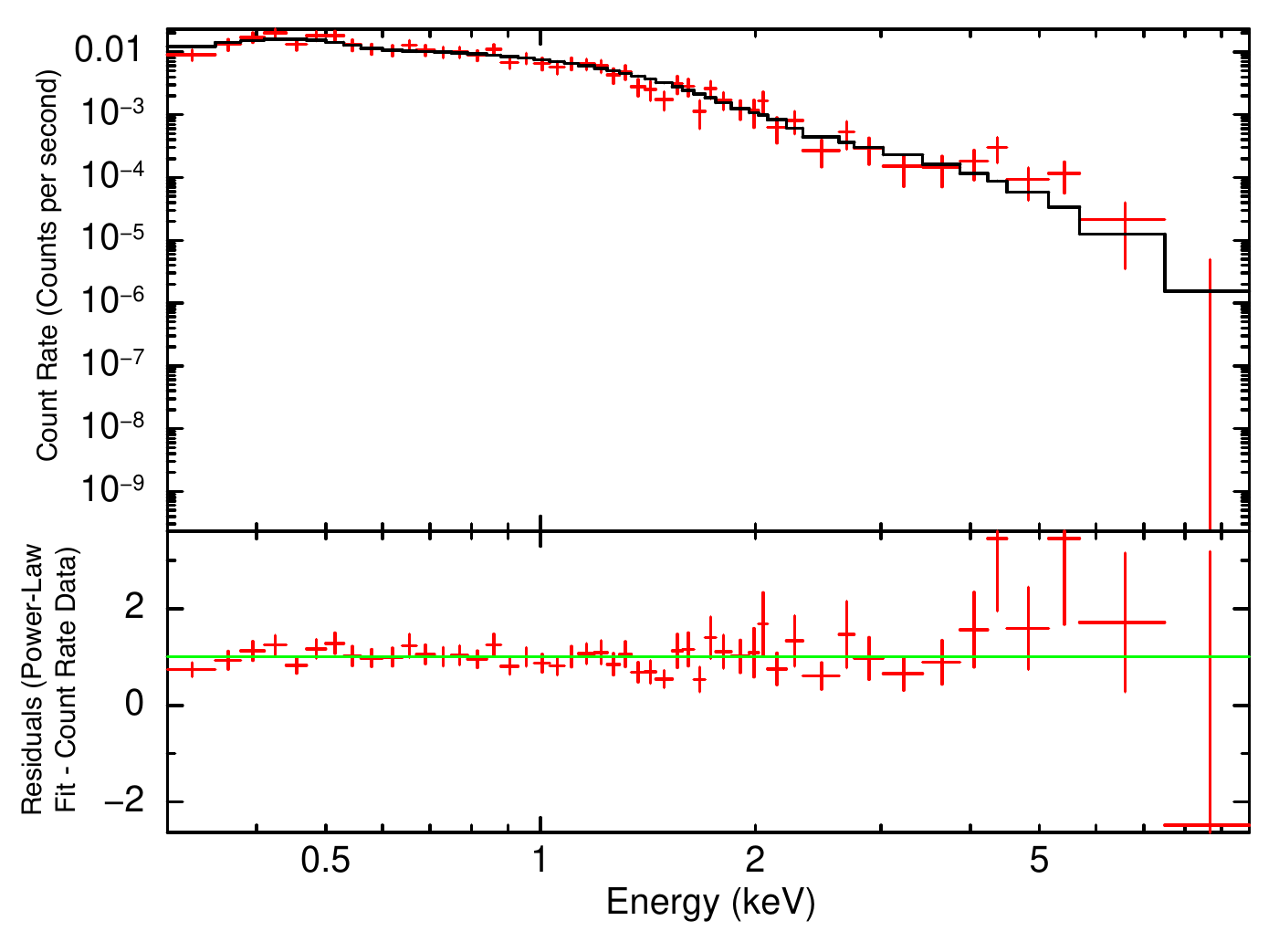}
    \caption{Top: Power-law fit to AT 2022upj during the ``pre-burst" phase (first 250 days from optical peak). The photon index for this phase is $\Gamma=2.55^{+1.37}_{-0.35}$. Bottom: Power-law fit to AT 2022upj during the ``burst" phase (after 250 days from optical peak) The photon index for this phase is $\Gamma=3.08^{+0.23}_{-0.21}$. }
    \label{fig:xrt_powerlaw}
\end{figure}

\section{Light Curve Model Fitting Outputs}
\label{sec:mosfit_outputs}

Here we provide the outputs from MOSFiT and TDEMass which estimate TDE parameters for AT\,2022upj from each model. Figure \ref{fig:mosfit_lc} shows the fit to the UV-optical light curve from MOSFiT, while Figure \ref{fig:mosfit_corner} shows the corner plot of fitted parameters: photospheric radius $R_\mathrm{ph0}$, viscous time delay $T_{\text{viscous}}$, scaled impact parameter $\beta$, mass of the supermassive black hole $M_{h}$, efficiency parameter $\epsilon$, photosphere power-law exponent $l$, host column density $n_\mathrm{H, host}$, mass of the disrupted star, time between first observation and disruption $t_\mathrm{exp}$, and white noise parameter $\sigma$ \citep{Mockler2019}. Figure \ref{fig:tdemass_solutions} shows the solution space for the mass of the SMBH $M_\mathrm{BH}$ and the disrupted star $M_\star$ from TDEMass \citep{Ryu2020}.

\begin{figure*}
    \centering
    \includegraphics[scale=0.6]{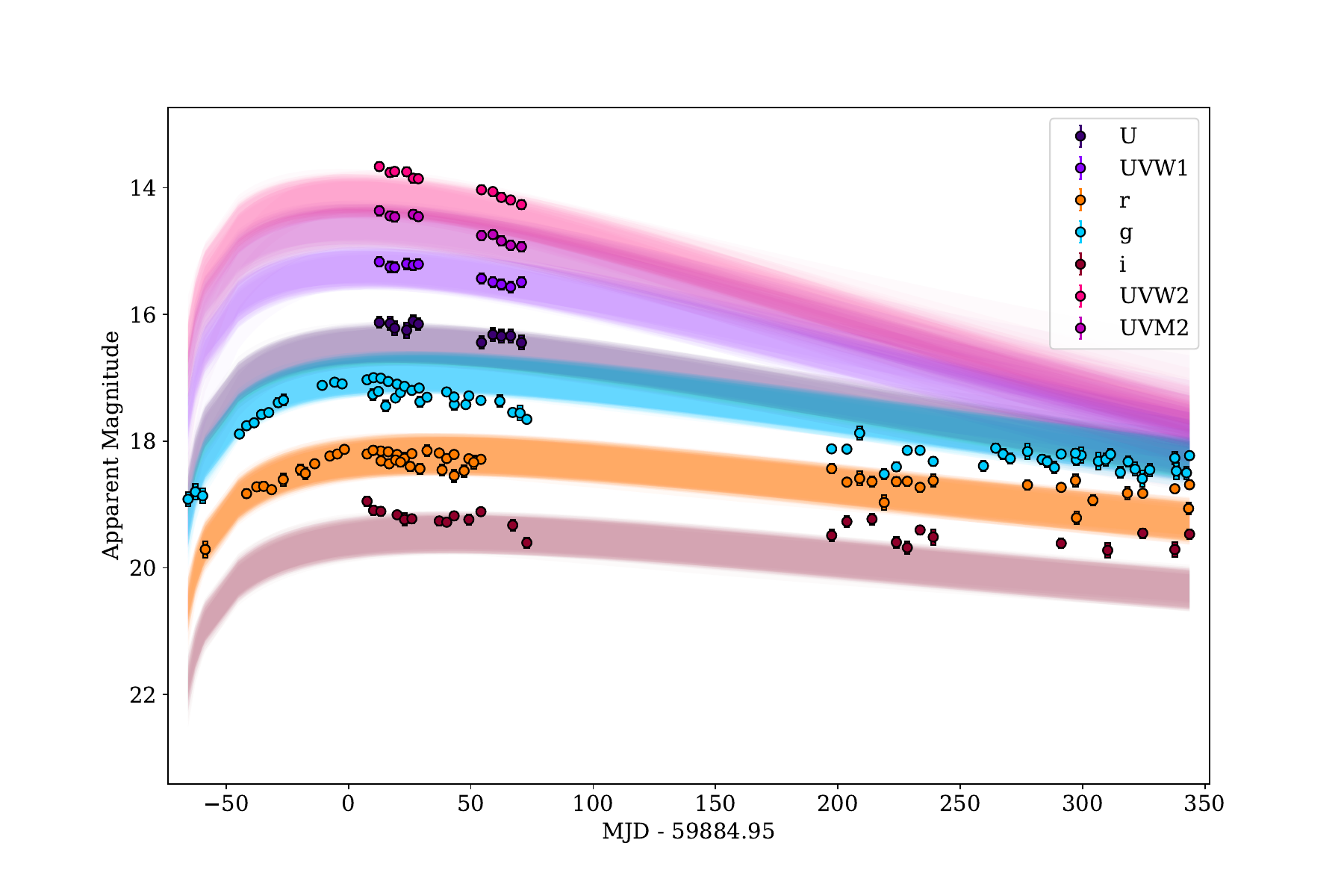}
    \caption{Light curve fitting output from MOSFiT to the multiband optical and UV photometry from Las Cumbres, ZTF, and Swift of AT\,2022upj.}
    \label{fig:mosfit_lc}
\end{figure*}

\begin{figure*}
    \centering
    \includegraphics[scale=0.3]{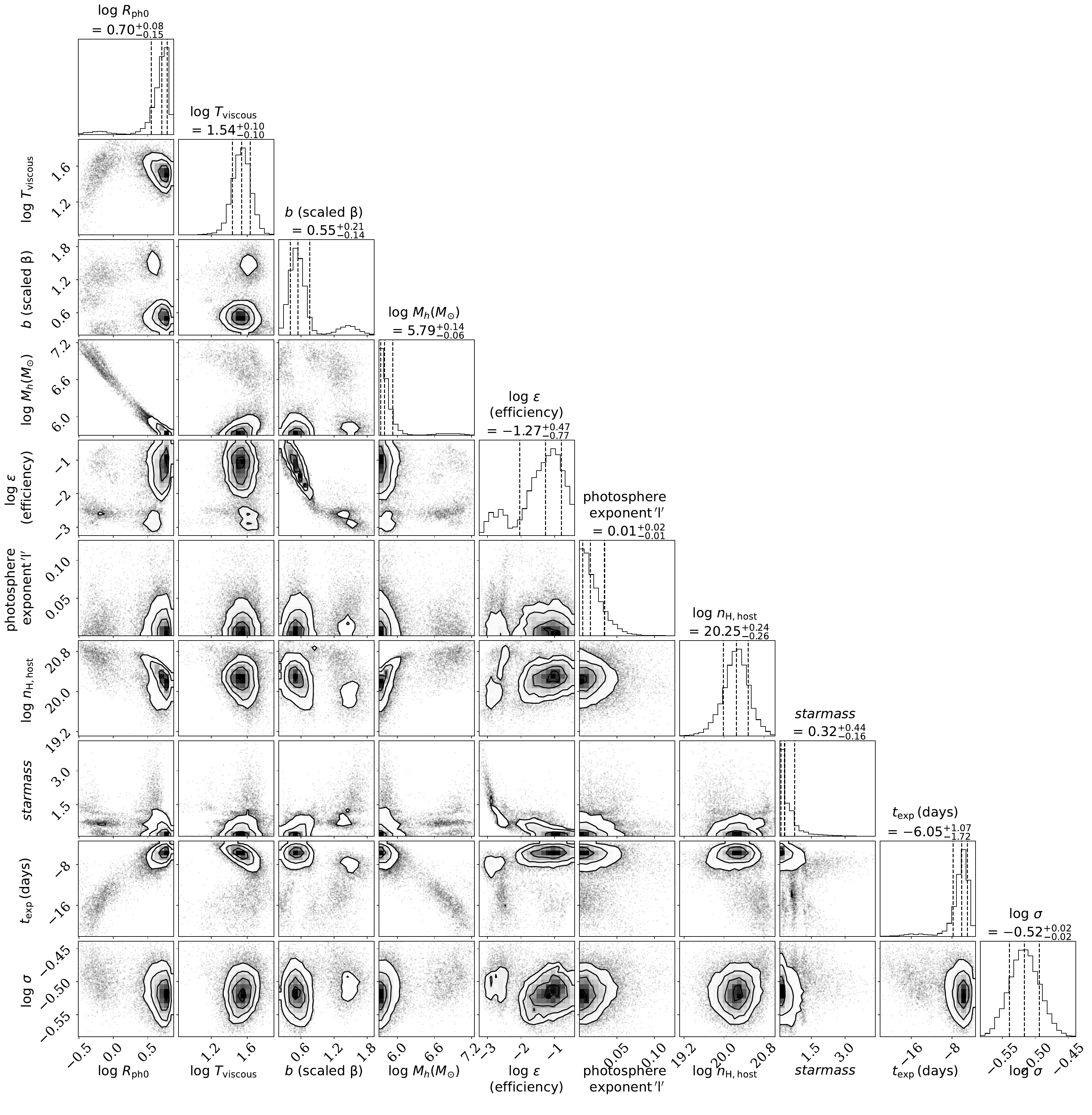}
    \caption{Corner plot of fitted parameters via nested sampling light curve fitting from MOSFiT.}
    \label{fig:mosfit_corner}
\end{figure*}

\begin{figure*}
    \centering
    \includegraphics[scale=0.5]{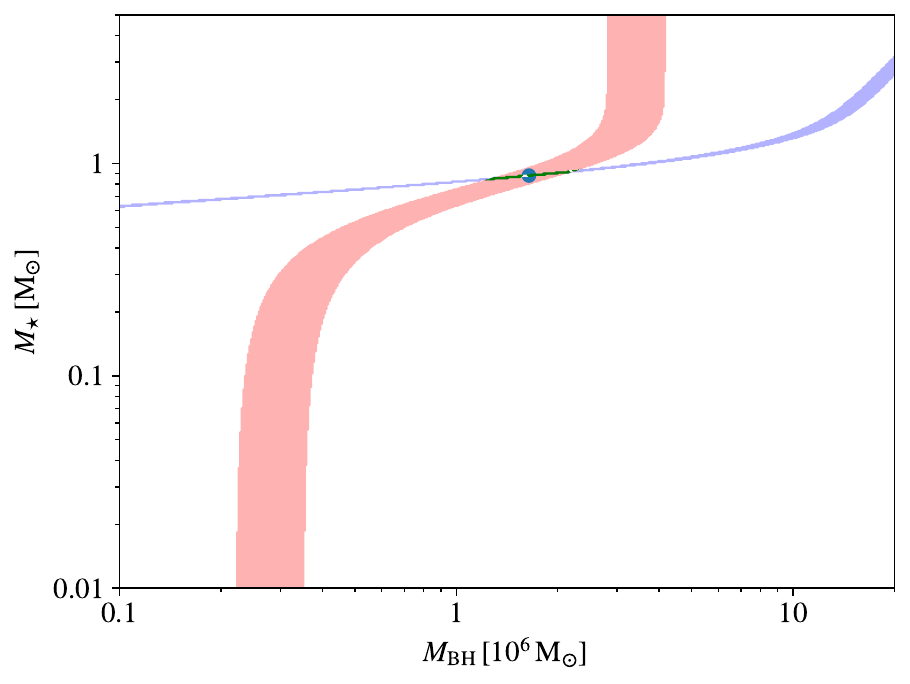}
    \caption{The solution for stellar mass and black hole mass from the peak temperature and bolometric luminosity of AT 2022upj as determined by TDEMass.}
    \label{fig:tdemass_solutions}
\end{figure*}

\section{Host Galaxy Fits}
\label{sec:host_fits}
\texttt{BAGPIPES} \citep{Carnall2018} fits spectroscopic models of galaxies to observed photometry in order to estimate the host's stellar mass and star formation history, metallicity, dust content, and age. We use the same choice of SFH and priors as in \cite{Carnall2018} for our initial fit. We use the stellar synthesis models as outlined in \cite{BC03} and updated in 2016\footnote{\url{http://www.bruzual.org/~gbruzual/bc03/Updated_version_2016/}} assuming a \cite{Chabrier2003} initial mass function. The star formation history is modeled as a double-power law with log-flat priors on the exponents in the power law ($\alpha$ and $\beta$ in Equation 10 of \citealt{Carnall2018}) in which we also use a uniform prior on metallicity and the \cite{Calzetti2000} dust law with a uniform prior on extinction. We fix the redshift to the spectroscopically determined value $z=0.054$. The host photometry used in these fits is listed in Table \ref{tab:hostphot} and the result of the spectroscopic fit is in Figure \ref{fig:bagpipes_specfit}.

\begin{figure*}
    \centering
    \includegraphics[scale=0.5]{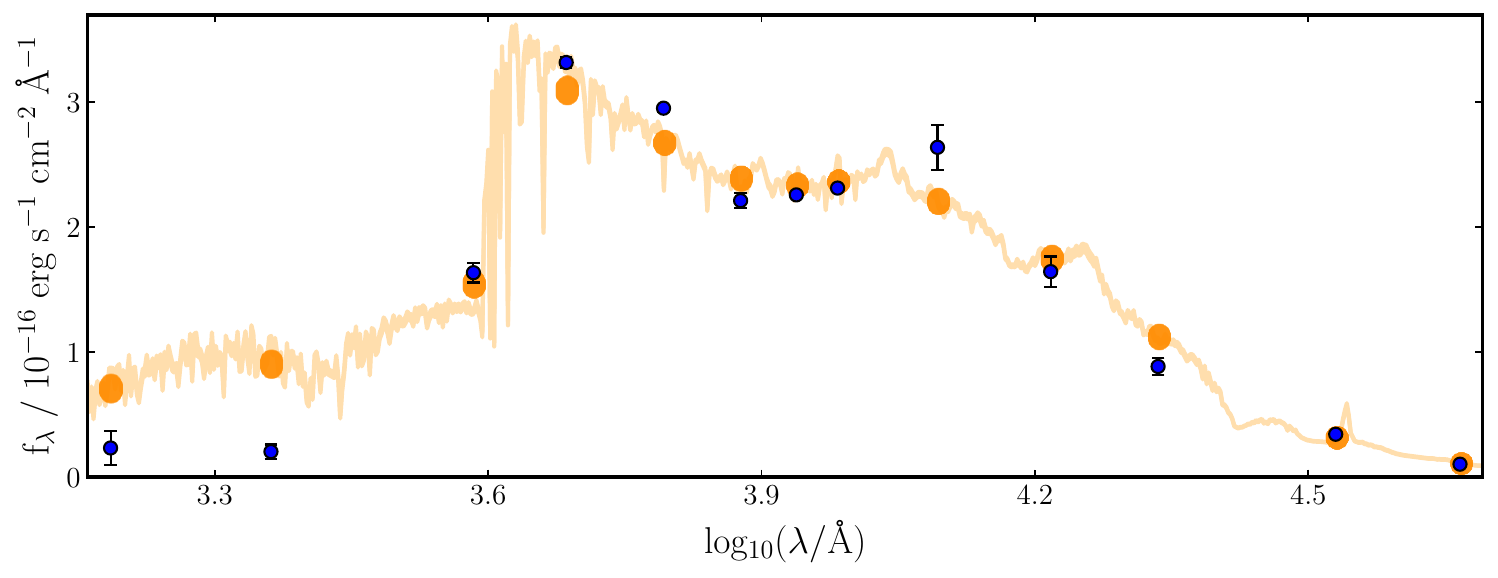}
    \caption{\texttt{BAGPIPES} spectral fit (orange line) and synthetic photometry (orange points) to the J003113.52+850031.8 photometrical SED (blue points) from the values in Table \ref{tab:hostphot}.}
    \label{fig:bagpipes_specfit}
\end{figure*}

\begin{figure*}
    \centering
    \includegraphics[scale=0.5]{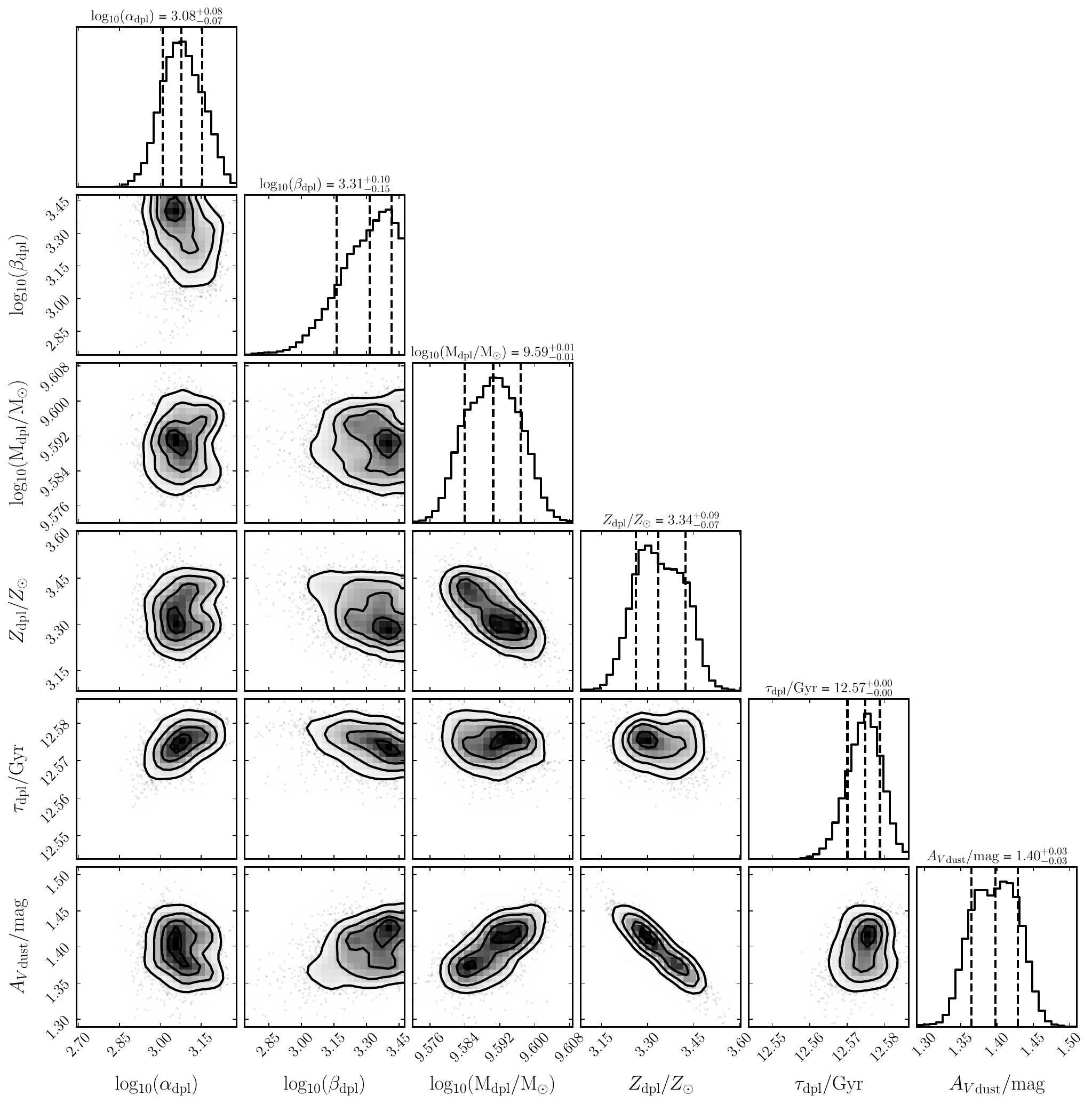}
    \caption{\texttt{BAGPIPES} corner plot of fitted parameters of the host galaxy of AT\,2022upj.}
    \label{fig:bagpipes_corner}
\end{figure*}

\begin{deluxetable}{cccc}[t!]
\label{tab:hostphot}
    \caption{Photometry of the host galaxy WISEA J002356.88-142523.9 used in fitting with \texttt{BAGPIPES}. All magnitudes are given in the Vega system except for $grizy$ magnitudes, given in the AB system.}
    \tablehead{
    \colhead{Filter} & \colhead{Source}& \colhead{Magnitude} & \colhead{Flux} \\[-6pt]
    \colhead{} & \colhead{} & \colhead{} & \colhead{$\mu$Jy}
    }
    \startdata
         FUV & GALEX & 20.19 & 1.82\\
         NUV & GALEX & 19.83 & 3.58\\
         $g$ & Pan-STARRS & 18.46 & 259.89\\
         $r$ & Pan-STARRS & 17.73 & 378.18\\
         $i$ & Pan-STARRS & 17.34 & 418.40\\
         $z$ & Pan-STARRS & 17.02 & 566.01\\
         $y$ & Pan-STARRS & 16.77 & 714.14\\
         $J$ & 2MASS & 15.54 & 1350.0\\
         $H$ & 2MASS & 14.37 & 1490.0\\
         $K$ & 2MASS & 14.47 & 1380.0\\
         $W1$ & WISE & 13.93 & 1310.0\\
         $W2$ & WISE & 13.89 & 736.0\\
    \enddata
\end{deluxetable}

\end{document}